\documentstyle[epsfig,12pt]{article}
\def\bc{\begin{center}}
\def\ec{\end{center}}
\def\be{\begin{equation}}
\def\ee{\end{equation}}
\def\bea{\begin{eqnarray}}
\def\eea{\end{eqnarray}}
\def\nn{\nonumber}

\def\simge{\ \lower-1.2pt\vbox{\hbox{\rlap{$>$}\lower5pt
\vbox{\hbox{$\sim$}}}}\ }
\begin{document}
\pagestyle{empty} 
\vspace{-0.6in}
\begin{flushright}
ROME1-1222/98 \\ 
\end{flushright}
\vskip 2cm
\centerline{\large{\bf{Heavy-heavy form factors and generalized
factorization}}}
\vskip 1.0cm
\centerline{M.~Ciuchini$^1$, R.~Contino$^2$, E.~Franco$^2$,
G. Martinelli$^{2}$}
\centerline{\small $^1$  Dipartimento di Fisica, Universit\`a di Roma Tre and INFN}
\centerline{\small Sezione di Roma III, Via della Vasca Navale 84,
I-00146 Rome, Italy.}
\centerline{\small $^2$ Dipartimento di Fisica, Universit\`a ``La Sapienza" and INFN,}
\centerline{\small Sezione di Roma, P.le A. Moro 2, I-00185 Rome, Italy.}
\vskip 2.0cm
\abstract{}
We reanalyze $B\to D\pi$ and  $B \to K J/\psi$  data to extract a set
of parameters which give the relevant hadronic matrix elements in terms
of factorized amplitudes. Various sources of theoretical uncertainties are
studied, in particular those depending on the model adopted  for the form factors.
We find that the fit to the $B\to D\pi$ branching ratios
substantially depends  on the model
describing the Isgur-Wise function and on the value of its slope.
This dependence can be  reduced by substituting the
$BR(B\to D\pi)$ with suitable ratios of
non-leptonic to differential semileptonic $BR$s.  In this way,
we obtain a model-independent determination of these parameters.
Using these results, the  $B\to D$ form factors
at $q^2=M_\pi^2$ can be extracted from a fit of  the $BR(B\to D\pi)$.
The comparison between the form factors obtained in this way
and the corresponding measurements in semileptonic decays  can be used
as a test of (generalized) factorization free from the uncertainties
due to heavy-heavy form factor modeling.
Finally, we present predictions for yet-unmeasured
$D\pi$ and $DK$ branching ratios and  extract $f_{D_s}$ and $f_{D_s^*}$ from
$B\to DD_s$ decays. We  find $f_{D_s}=270\pm 45$ MeV and
$f_{D_s^*}=260\pm 40$ MeV, in good  agreement with recent measurements and
lattice calculations.
\vfill
\begin{flushleft} 
ROME1-1222/98 \\
Oct. 1998
\end{flushleft}
\eject
\pagestyle{empty}\clearpage
\setcounter{page}{1}
\pagestyle{plain}
\newpage 
\pagestyle{plain} \setcounter{page}{1}

\section{Introduction}
\label{sec:introduction} 
A problem of utmost importance in $B$ phenomenology is the computation of
the hadronic amplitudes: in recent years it has been realized that the
full determination of the unitarity triangle from $B$ decays can hardly
be carried out without an accurate knowledge of these quantities~\cite{chp,cprev}.
Unfortunately, the computation of hadronic amplitudes requires an understanding of low-energy
strong interactions which is missing at present. Even a non-perturbative
approach based on first principles, such as lattice QCD, fails in computing
decay amplitudes involving two or more hadrons in the final state~\cite{mt}.

In the absence of rigorous methods, some simplifying approaches have been
developed. The most popular one consists in the factorization of matrix
elements of four-fermion operators  in terms of local products of two currents.
In this approach, the original matrix element is computed as  product of
the matrix elements of the two currents.
Attempts to give theoretical soundness to this procedure in the framework of
the $1/N$ expansion and of the Large Energy Effective Theory (LEET) can be found in
refs.~\cite{bgr,dg}.
Unfortunately, there are many problems in both these approaches
and their applicability to exclusive decays is questionable.
Independently of any theoretical prejudice, 
there is a priori {\it no} reason for this
approximation to be accurate in the case of $B$ decays.
Indeed, none of the expansions developed so far was able to
compute corrections to the lowest order results and estimate the size of the errors. 
On the other hand, the
importance of controlling the theoretical uncertainties calls for some
phenomenological approach to test predictions obtained using
factorized amplitudes. 
To this end, a popular method consists in reducing the Wick contractions of
matrix elements to few topologies using Fierz transformations and
color rearrangement. Then, the remaining amplitudes are factorized and
expressed in terms of the appropriate decay constants and/or form factors.
In this procedure, some phenomenological parameters are introduced
in order to account for possible deviations from
factorization~\cite{chp,bsw87,neuste}.
These factorization parameters, denoted as FP in the following,
are meant to be fitted to experimental data.

In this paper, we introduce a parameterization of the hadronic matrix elements
that extends the one of ref.~\cite{chp} 
and allows the computation of
the hadronic amplitudes relevant to Cabibbo-allowed non-leptonic $B$ decays in
terms of factorized matrix elements and of three real FP.
We find that, in the fit of the  $BR(B\to D\pi)$ 
and $BR(B\to K J/\Psi)$,   there  is a strong interplay
between the values of the FP and  the model used for the 
 heavy-heavy form factors, more specifically  on the Isgur-Wise (IW)
function and its slope $\rho^2$~\footnote{ Here and in the following,
unless stated explicitly otherwise , $B \to D \pi$ denotes generically  a full set of  decays of
 $B$ mesons into a charmed and a light meson, i.e. $B_d \to \pi^+ D^-$, 
$B_d \to \rho^+ D^-$, $B^+ \to \pi^+ \bar D^{*0}$, etc.}.
This implies that factorization  tests are
obscured by our ignorance  on the values of the form factors in the
kinematical region relevant in non-leptonic decays ($q^2 \ll q^2_{max}$).
The model dependence is drastically reduced by
using,  in the fit, suitable ratios of semileptonic and non-leptonic $BR$s
(to be introduced below)  instead of the  $BR(B\to D\pi)$ alone. 
In this way, we are able to extract (almost) model independent FP.
With these FP at hand, we use then the $BR(B\to D\pi)$ to determine, within a given
model,  the value of $\rho^2$ which may be compared, as a test of factorization,
to the one measured in semileptonic decays. 
The value of $\rho^2$ extracted from the fit depends, however, on the
model used for the form factors. Different values of $\rho^2$ compensate, indeed, for the
different dependence of the theoretical form factors on the momentum transfer, 
thus giving the same values for the matrix elements of the weak currents 
at low $q^2$. 
We conclude that the quantities to be compared with the corresponding ones 
in semileptonic decays 
are the form factors themselves  in the region of $q^2$ relevant in non-leptonic decays 
(e.g. $q^2=M_\pi^2\sim 0$ for $B \to D \pi$ decays).  This is  a real test of
factorization, free from model uncertainties.  The values
of the form factors extracted from our analysis are given in table~\ref{tab:ff}.
\begin{table}[t]
\centering
\begin{tabular}{|c|c|c|}
\hline
Form Factor & LINSR & NRSX \\
\hline
$f_+(M_\pi^2)$  & 0.56 [0.49--0.63] & 0.57 [0.52--0.63] \\
$V(M_\pi^2)$    & 0.68 [0.61--0.73] & 0.75 [0.69--0.81] \\
$A_0(M_\pi^2)$  & 0.59 [0.54--0.64] & 0.58 [0.54--0.63] \\
$A_1(M_\pi^2)$  & 0.59 [0.53--0.64] & 0.56 [0.52--0.61] \\
$A_2(M_\pi^2)$  & 0.59 [0.53--0.64] & 0.54 [0.49--0.58] \\
\hline
\end{tabular}
\label{tab:ff}
\caption{\it{Values of $B\to D$ form factors determined by fitting
the $B\to D\pi$ data with two different models, as explained in the text below. 
The ranges in square brackets correspond to variations
of $\chi^2/dof$ up three times larger than its  minimum.
We find that form factors at $q^2=M_\pi^2$ already verify the
kinematical relation $f_0=f_+$ and $A_3=A_0$ valid at $q^2=0$.
}}
\end{table}
%
%
In principle, one may extract the five form factors of tab.~\ref{tab:ff}
independently. However, in our analysis, all the form factors are related
to the IW function through the heavy quark symmetry.
Consequently, the only $B\to D$ form factor measured so far at small $q^2$,
$f_+(0)$, already allows a full test of our approach.
Its experimental value, $f_+(0)=0.66 \pm 0.06 \pm 0.04$~\cite{cleofp}
is in good agreement with our findings. Measurements of
the other form factors entering $B \to D^*$ semileptonic decays
would check the relations enforced by the heavy quark symmetry.

The  determination of the FP also allows us  to predict several $BR$s,  
including $B\to DK$, which have  not been measured yet.
Our predictions are presented in table~\ref{tab:pred}.

Finally, using ratios of  non-leptonic $BR$s involving
$D^{(*)}D_s^{(*)}$ final states and the FP and form factors from  the previous fits, we extract the
meson decay constants $f_{D_s}$ and $f_{D_s^*}$, obtaining
\be
f_{D_s}=270\pm 45~\mbox{MeV},\qquad f_{D_s^*}=260\pm 40~\mbox{MeV},
\ee
in good agreement with recent experimental measurements $f_{D_s}=250 \pm 30$
MeV~\cite{nippe}  and lattice determinations, $f_{D_s}
= 218^{+20}_{-14}$ MeV (quenched), $f_{D_s}
= 235^{+22+17}_{-15-9}$ MeV (unquenched)~\cite{fdslat} and $f_{D_s^*}= 240 \pm 20$ MeV 
(preliminary quenched)~\cite{fdsslat}.
We also study the contribution of charming penguins~\cite{chp} and discuss
their effects on the predictions for the decay constants,
which we find non-negligible.

The paper is organized as follows. In sect.~\ref{sec:factorization} we
introduce our FP and two different  models for the form factors to be used in
the phenomenological analysis. Section~\ref{sec:fit} contains the
main results of our fits to the $B\to D\pi$ and $B\to KJ/\Psi$ branching ratios, namely
the determination of the FP, the analysis of their $\rho^2$ dependence and
the extraction of the $B\to D$  and $B \to D^*$ form factors at $q^2=M_\pi^2$. 
The results of these fits have been used for the predictions of yet-unmeasured $BR$s, 
including many $B\to DK$ modes.
Finally, in sect.~\ref{sec:costdec}, we analyze the $B\to DD_s$ modes and
extract $f_{D_s}$ and $f_{D_s^*}$, giving an estimate of the
theoretical error which includes charming-penguin effects.

\begin{table}[t]
\centering
\begin{tabular}{|l|cr|cr|c|}
\hline
Channel & \multicolumn{2}{c|}{LINSR} & \multicolumn{2}{c|}{NRSX} & Experiment \\
	& \multicolumn{2}{c|}{$[BR\times 10^5]$} &
          \multicolumn{2}{c|}{$[BR\times 10^5]$} &  $[BR\times 10^5]$ \\
\hline
$B_d\to \pi^0 \bar D^0$       & 14 & [1--58]  & 10 & [2--27]  & $<12$ \\
$B_d\to \pi^0 \bar D^{*0}$    & 15 & [1--63] & 13 & [2--35] & $<44$ \\
$B_d\to \rho^0 \bar D^0$     & 6  & [1--26]  & 7  & [1--19]  & $<39$ \\
$B_d\to \rho^0 \bar D^{*0}$  & 17 & [2--71] & 14 & [2--38] & $<56$ \\
\hline
$B_d\to K^+ D^-$	     & 22 & [13--38] & 23 & [16--32] & -- \\
$B_d\to K^+ D^{*-}$	     & 22 & [14--37] & 22 & [16--30] & -- \\
$B_d\to K^{*+} D^-$	     & 53 & [32--90] & 53 & [38--75] & -- \\
$B_d\to K^{*+} D^{*-}$	     & 67 & [43--110]& 64 & [46--88] & -- \\

$B^+\to K^+ \bar D^0$	     & 35 & [12--54] & 35 & [18--45] & $29 \pm 10$ \\
$B^+\to K^+ \bar D^{*0}$     & 36 & [13--54] & 35 & [17--44] & -- \\
$B^+\to K^{*+} \bar D^0$     & 67 & [34--100] & 69 & [42--88] & -- \\
$B^+\to K^{*+} \bar D^{*0}$  & 87 & [46--126]& 83 & [51--104]& -- \\

$B_d\to K_0 \bar D^0$	     & 1.4 & [0.2--5.7] & 1.0 & [0.2--2.7] & -- \\
$B_d\to K_0 \bar D^{*0}$     & 1.5 & [0.2--6.2]  & 1.3 & [0.2--3.4] & -- \\
$B_d\to K^{*0} \bar D^0$     & 0.6 & [0.1--2.7]	& 0.7 & [0.1--1.8] & -- \\
$B_d\to K^{*0} \bar D^{*0}$  & 1.7 & [0.2--7.2]	& 1.4 & [0.2--3.6] & -- \\
\hline

\end{tabular}
\caption{{\it Predictions of yet-unmeasured branching ratios.}}
\label{tab:pred}
\end{table}

\section{Factorization, FP and form-factor models}
\label{sec:factorization} 

In this section, we present our parameterization of the hadronic
amplitudes and discuss its relation with other popular choices.
We also introduce two different models for the form factors 
used in our phenomenological analysis.

Consider the matrix element of some composite operator appearing in the
$\Delta B=1$ weak Hamiltonian, between the $B$ meson and two final 
pseudoscalar or vector mesons.
In general this operator can be written as the product of two currents.
If one of the currents has the correct quantum numbers to create  one
of the final state mesons  from the vacuum, then the matrix element can be factorized. 
The physical idea
is the following: the quark pair produced by this current acts as a
color dipole, weakly interacting with the surrounding color field.
If the transferred energy is large, the quark pair has no time to interact
before hadronizing far from the interaction point~\cite{bj}. 

As an example, we discuss the factorization of the amplitudes entering
the decay $B_d\to D^-\pi^+$.  In this case, the two relevant matrix elements
($\alpha$ and $\beta$ are color indices)
\bea
\langle D^-\pi^+\vert Q_1\vert B_d\rangle &=&
\langle D^-\pi^+\vert \bar b_\alpha \gamma_\mu (1-\gamma_5) c_\beta
~\bar u_\beta \gamma^\mu (1-\gamma_5) d_\alpha \vert B_d\rangle ,\nonumber \\
\langle D^-\pi^+\vert Q_2\vert B_d\rangle &=&
\langle D^-\pi^+\vert \bar b_\alpha \gamma_\mu (1-\gamma_5) c_\alpha
~\bar u_\beta \gamma^\mu (1-\gamma_5) d_\beta \vert B_d\rangle ,
\label{eq:oper}
\eea
can be Wick-contracted according to two different topologies,
that are usually denoted as connected ($CE$) and disconnected ($DE$)
emissions,  respectively.
Color indices can be rearranged using the algebraic relation,
\be
\delta_{\alpha\beta}\delta_{\rho\sigma}=\frac{1}{N}\delta_{\alpha
\sigma}\delta_{\rho\beta}+2t^a_{\alpha\sigma}t^a_{\rho\beta},
\ee
where $N$ is the number of colors, $\delta$ is the Kronecker symbol and the $t^a$
are the $SU(N)$ color matrices in the fundamental representation,
normalized  as $tr(t^a t^b)=\delta^{ab}/2$.
Using this relation, one obtains
\\ \vspace{0.5cm} \\
{\large
\hspace*{0.7cm}
 \begin{minipage}[c]{.20\linewidth}
 \vspace*{0.33cm}
 \centering \epsfig{file=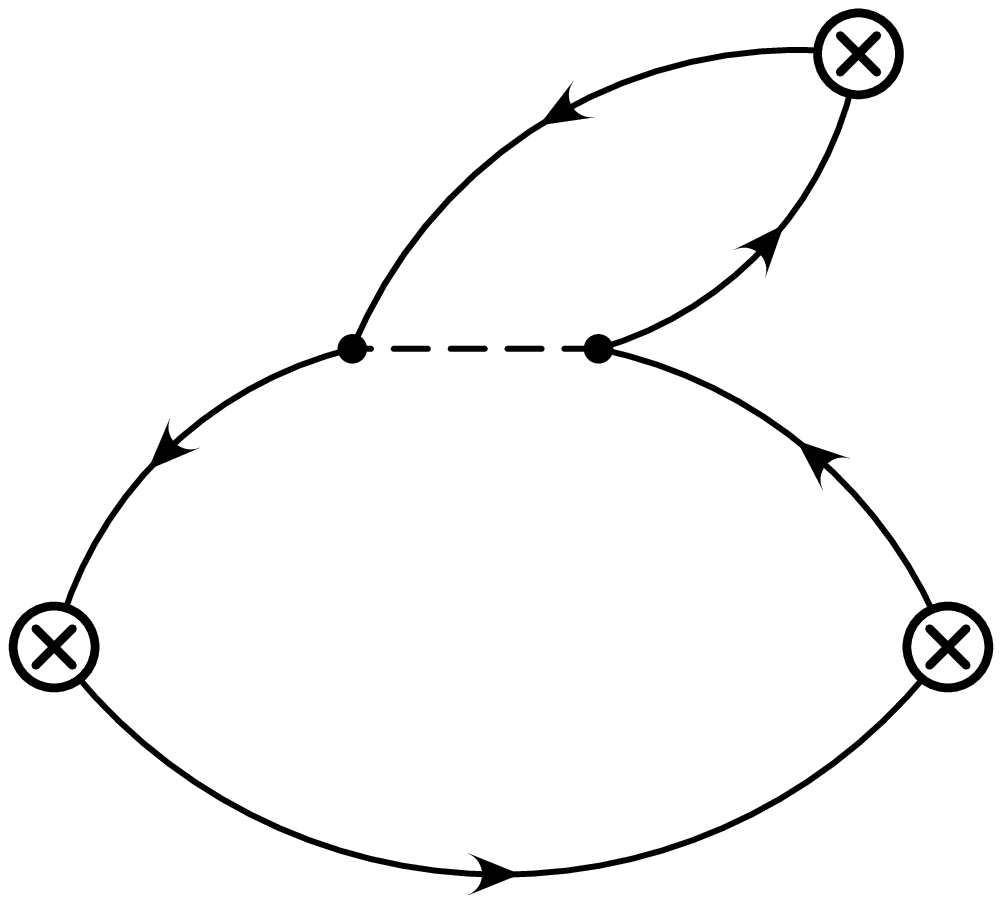,width=\linewidth}\\
 \centering CE
\end{minipage}
$=\frac{1}{N}$
\begin{minipage}[c]{.20\linewidth}
 \centering \epsfig{file=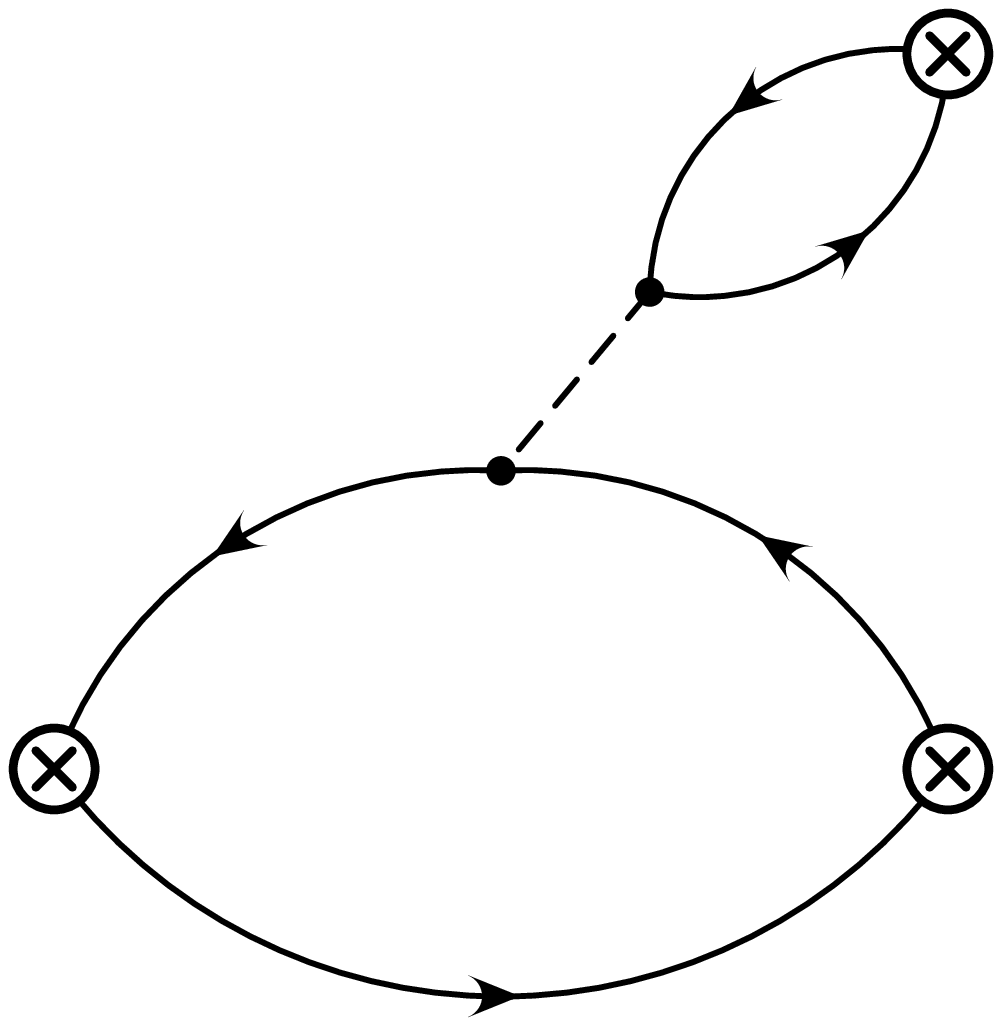,width=\linewidth}\\
 \centering DE
\end{minipage}
$+2$
\begin{minipage}[c]{.20\linewidth}
 \centering \epsfig{file=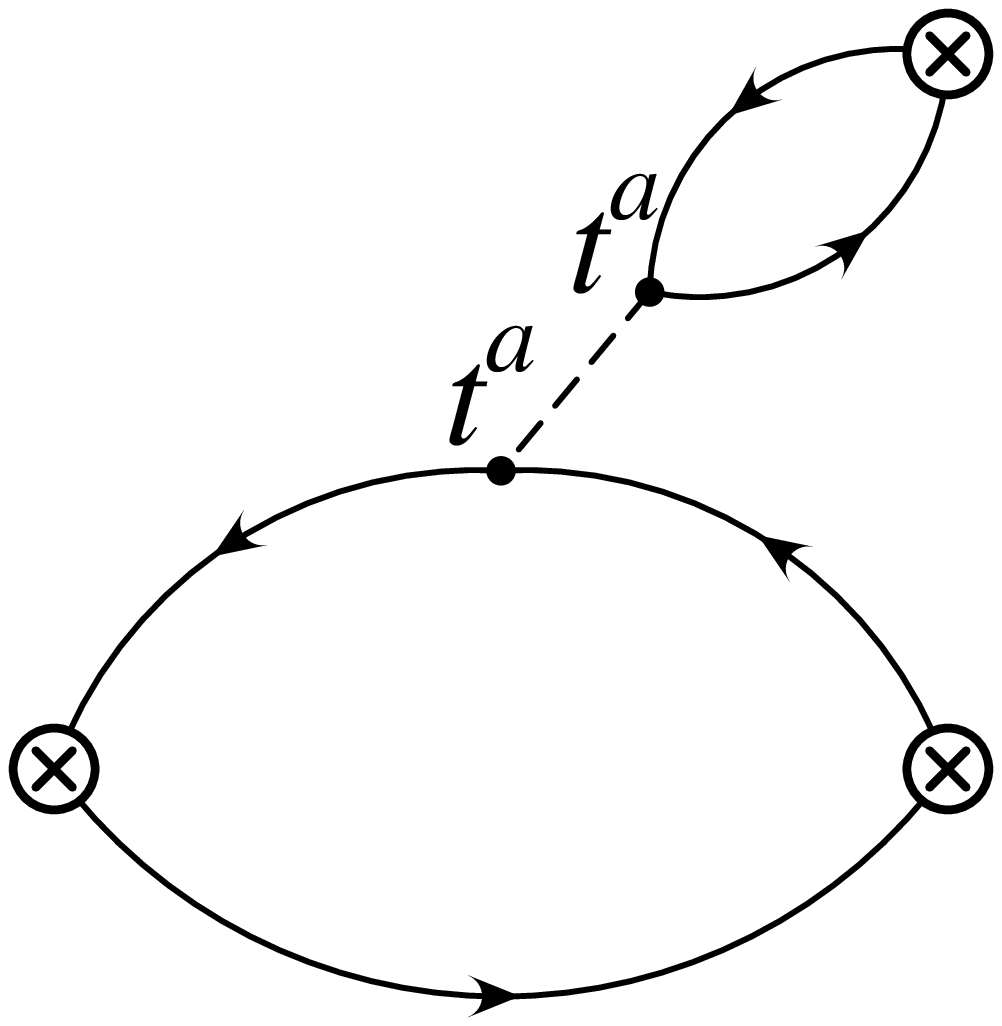,width=\linewidth}\\
 \centering octet terms
\end{minipage}~.
}
\\ \vspace{0.5cm} \\
In the factorization limit, no gluon exchange occurs between the quark
pair of the emitted meson and the other quarks,  so that the octet terms 
vanish  and the relation between $DE$ and $CE$ becomes simply
$CE=DE/N$. 
Exact factorization is known to fail, however,  in
reproducing $D$ phenomenology~\cite{dphen}. For this reason,
it is customary to introduce several phenomenological parameters 
to account for octet terms (and in general
for the different  sources of factorization
violation). These parameters may be extracted from the experimental data.
An example is provided by the generalized factorization of ref.~\cite{neuste}. In this
case  the relevant contractions are rewritten, without loss of generality, as
\be
DE=(1+\epsilon_1)DE_{fact}\ ,\qquad CE=\left(\frac{1}{N}+\frac{\epsilon_8}
{1+\epsilon_1}\right) (1+\epsilon_1) DE_{fact} \ , 
\label{eq:genfact}
\ee
where the two parameters, $\epsilon_1$ and $\epsilon_8$, vanish in the
factorization limit. 

 In this paper, following ref.~\cite{chp}, we adopt a different parameterization, 
given by
\be
DE=\alpha DE_{fact},\qquad CE=\alpha e^{i\delta_\xi}\xi DE_{fact},
\label{eq:genfact2}
\ee
where $DE$ and $CE$ are given in terms of three real parameters $\alpha$, $\xi$ and
$\delta_\xi$. 
Note that there is no inconsistency between
the two parameterizations: in general, there are three  real parameters,
namely two moduli $\vert DE\vert$, $\vert CE\vert$, and one 
relative phase, $\arg(CE)-\arg(DE)$. These correspond to our three real
parameters $\alpha$, $\xi$ and $\delta_\xi$ or to the two complex
parameters $\epsilon_1$ and $\epsilon_8$ in eq.~(\ref{eq:genfact}),
one of which can always be chosen real. 
The relation between the two sets of parameters is given by
\be
\alpha=1+\epsilon_1,\qquad \xi e^{i\delta_\xi}=\frac{1}{N}+\frac{\epsilon_8}
{1+\epsilon_1}.
\ee

As recently stressed in ref.~\cite{busil}, these phenomenological parameters
are renormalization scale and scheme dependent, as much as
the original matrix elements, since the factorized amplitudes are insensitive
to both the scale and the scheme. This dependence is required to cancel the
corresponding dependence in the Wilson coefficients, up to the order at which
the perturbative calculation is done. Note that,  in order to study the scale
dependence  of the parameters, the next-to-leading order (NLO)
determination of the effective Hamiltonian is required.
Being scheme-dependent, any physical interpretation
of the ``factorization scale'', namely of the renormalization scale (if it
really exists) at which exact factorization holds, is meaningless.
Nevertheless, the FP can be precisely
extracted from data, once the  renormalization scale and the scheme are fixed. 
Their values will depend, of course,  on these choices.
We will use the NDR-$\overline{\rm MS}$ NLO Wilson
coefficients computed at
$\mu=5$ GeV, as given in ref.~\cite{zeitnoi}.
In the following, it is understood that we determine $\alpha$, $\xi$ and
$\delta_\xi$ using this choice of the scale and of the renormalization scheme.

After the introduction of the FP, the only amplitude that
remains to be computed, namely
\bea
DE_{fact}&=&\langle D^-\pi^+\vert \bar b \gamma_\mu (1-\gamma_5) c
~\bar u \gamma^\mu (1-\gamma_5) d \vert B_d\rangle\vert_{fact}\nonumber\\
&=&\langle D^-\vert \bar b \gamma_\mu (1-\gamma_5) c
\vert B_d\rangle\langle \pi^+\vert
\bar u \gamma^\mu (1-\gamma_5) d \vert 0\rangle~,
\eea
can be easily expressed in terms of the $B\to D$ semileptonic form factors and 
of the decay constant $f_\pi$.

In this example, only left-handed currents appear. In general, considering
also penguin operators, there are 
diagrams involving different Dirac structures, {\it e.g.}
$\gamma^\mu (1-\gamma_5)\otimes \gamma^\mu (1+\gamma_5)$ and
$(1-\gamma_5)\otimes (1+\gamma_5)$.
In the case of interest, the right-handed current always appears in the matrix
element of the emitted meson, $\langle M\vert
\bar q \gamma^\mu (1+\gamma_5) q^\prime \vert 0\rangle$, while the current
entering the other matrix element is always left-handed.
Therefore only the vector or the axial current separately contributes, depending
on the quantum numbers of the emitted meson. 
Consequently, assuming that both left-left and left-right operators can be described with
the same set of FP, the relation between the corresponding matrix elements
becomes trivial. Similarly,
matrix elements of operators with a $(1-\gamma_5)\otimes (1+\gamma_5)$
Dirac structure can be connected
to the current-current ones via the vector and axial vector Ward identities~\footnote{
In this case the amplitudes depend on the quark masses which we take to be defined in the
same renormalization scheme, and at the same scale $\mu$, as the four-fermion operators.}.
In summary, using factorization, one only needs to compute matrix elements
of currents, that can be expressed in terms of form factors and/or decay
constants.
However, these relations among insertions of different Dirac structures
only hold for {\it factorized} amplitudes. By using only
one set of FP, we implicitly assume that the same relations hold for
the original four-fermion operator matrix elements.
This simplifying assumption allows us to 
account for penguin-operator contributions using factorization, albeit in a
model-dependent way.

In our analysis, we use the same FP for
{\it i}) matrix elements connected by $SU(2)$ flavor symmetry;
{\it ii}) matrix elements with the same quark content, differing only for
the angular momentum of the final hadrons.
The first assumption has sound phenomenological motivations;
the second is reasonable since some of the differences among matrix elements
with pseudoscalar and/or vector meson final states are already accounted for
by factorized matrix elements.

For a  generic transition $B\to P(V)$  of a $B$ going into a pseudoscalar (vector) meson
of mass $M$, momentum $p$ (and polarization $\epsilon$),  
the form factors in the helicity basis are defined as
\bea
\langle P(p)|\hat V_\mu|B(p_B)\rangle &=&
 f_+(q^2) \Bigg\{ (p+p_B)_\mu -q_\mu \frac{M_B^2-M^2}{q^2} \Bigg\} + 
 f_0(q^2) q_\mu \frac{(M_B^2-M^2)}{q^2}, \nn \\
\langle V(p,\epsilon)|\hat V_\mu|B(p_B)\rangle &=&
 \frac{2i}{M_B+M} \varepsilon_{\mu \nu \alpha \beta} \epsilon^{* \nu} p^\alpha 
 p_B^\beta V(q^2), \label{eq:Avett} \\
\langle V(p,\epsilon)|\hat A_\mu|B(p_B)\rangle &=&
 2 M A_0(q^2) q_\mu \frac{\epsilon^* \cdot q}{q^2} + A_1(q^2) (M_B+M) \Bigg\{
 \epsilon^*_\mu - q_\mu \frac{\epsilon^* \cdot q}{q^2} \Bigg\} \nn\\
 &&\quad -\frac{A_2(q^2)}{M_B+M} \Bigg\{ (p_B+p)_\mu \epsilon^* \cdot q - 
 q_\mu \frac{\epsilon^* \cdot q}{q^2} (M_B^2-M^2) \Bigg\},\nn
\eea 
where $\hat V_\mu$ and $\hat A_\mu$ are the vector and axial currents
respectively. 

For heavy final mesons, the form factors  can be connected to the HQET functions
$\xi_i(y)$, see {\it e.g.} ref.~\cite{neuri}, using the following formulas
\bea
\langle P(v)|\hat V_\mu|B(v_B)\rangle &=& \sqrt{M_B M}\;\Big\{
\xi_+(y) (v_B+v)_\mu + \xi_-(y)(v_B-v)_\mu\Big\}, \nn\\
\langle V(v,\epsilon)|\hat V_\mu|B(v_B)\rangle &=&\sqrt{M_B M}\;\Big\{
 i \xi_V(y) \varepsilon_{\mu \nu \alpha \beta} \epsilon^{* \nu}
 v^{\alpha} v_B^\beta\Big\}, \label{eq:VvettHQET}\\
\langle V(v,\epsilon)|\hat A_\mu|B(v_B)\rangle &=&\sqrt{M_B M}\;\Big\{
 \xi_{A_1}(y)(y+1) \epsilon^*_\mu \nn\\
 &&\quad -\xi_{A_2}(y) (\epsilon^* \cdot v_B) v_{B \, \mu}
 -\xi_{A_3}(y) (\epsilon^* \cdot v_B) v_{\mu}\Big\},\nn
\eea
where $v_B$ and  $v$ are the 4-velocities of the initial and final meson
respectively and 
\be
y \equiv v_B \cdot v = \frac{M_B^2 +M^2 -q^2}{2M_B M}.
\ee

In the heavy quark limit, the functions $\xi_i(y)$ are all related to
the IW function $\xi(y)$
\be
\xi_+ = \xi_{V} = \xi_{A_1} = \xi_{A_3} = \xi(y),\qquad
\xi_- = \xi_{A_2} = 0,\label{eq:limiteHQS}
\ee
with the normalization of $\xi$ fixed by the heavy quark symmetry, $\xi(1)=1$.
The $\xi_i(y)$ can be written as
\be
\xi_i(y) = \Big\{ \alpha_i +\frac{\alpha_s(\bar m)}{\pi} \beta_i(y) +
\gamma_i(y) \Big\}\,\xi(y),\label{eq:hhff}
\label{eq:hqetff}
\ee
where the functions $\beta_i(y)$ and $\gamma_i(y)$ take into account
the perturbative ${\cal O}(\alpha_s)$ corrections and the ${\cal O}(1/m)$ terms respectively
\footnote{The computation of the $1/m$ corrections is model dependent, relying
on the evaluation of a set of hadronic matrix elements of higher dimensional
operators in the HQET.}.  Following ref.~\cite{neuqcd}, we used for $\bar m$ the
reduced charm-bottom mass i.e.  $\bar m= 2.26$ GeV.

We are now ready to introduce the two models that we will use in order to study
the form-factor dependence of the FP. We denote these
models as LINSR and NRSX:
\begin{itemize} \item \underline{LINSR}: 
the first model uses the heavy-heavy form factors defined in 
eq.~(\ref{eq:hhff}), taking the $\beta_i(y)$ from ref.~\cite{neuqcd} and
neglecting the ${\cal O}(1/m)$ corrections, {\it i.e.} $\gamma_i(y)=0$.
For the IW function, the simplest form is assumed, namely
\be
\xi(y)= 1- \rho^2 (y-1)\ .
\label{eq:LINSRIW}
\ee
For the heavy-light form factors,  LINSR uses those computed with light-cone QCD sum
rules~\cite{qcdsr}. 
\item \underline{NRSX}:
the second model is the one defined in ref.~\cite{nrsx} and makes use of the functions
$\beta_i(y)$ and $\gamma_i(y)$  calculated in refs.~\cite{neuri} and
\cite{neuqcd}, respectively.
The IW function is obtained using a relativistic oscillator model, which gives
\be
\xi(y) = \frac{2}{y+1} \exp \bigg\{ -A \; \frac{y-1}{y+1} \bigg\},
\label{eq:NRSXIW}
\ee
where $A=2\rho^2 -1$. 
Concerning the heavy-light form factors, NRSX improves the old WSB model~\cite{wsb} by
implementing,   for $q^2 \sim q^2_{max}$, the expected heavy quark scaling laws, see ref.~\cite{nrsx}
for details.
\end{itemize}

We end this section by spending a word of comment about the parameter $\rho^2$
entering eqs.~(\ref{eq:LINSRIW}) and (\ref{eq:NRSXIW}).
In both models, $\rho^2$ is defined as the slope of the
IW function at the zero-recoil point ({\it i.e.} $y\equiv v_B\cdot v=1$),
namely
\be
\rho^2 \equiv - \frac{d}{dy} \xi(y) \bigg|_{\, y=1}.
\ee
The value of $\rho^2$ is related to the semileptonic 
differential rate for $B\to D^* l \nu$,
\bea
\frac{d\Gamma(B\to D^* l \nu)}{dy} &=& \frac{G_F^2}{48\pi^3} M_{D^*}^3 (M_B-M_{D^*})^2
 \sqrt{y^2-1} (y+1)^2 \nn\\
 &&  \times  \bigg[ 1+\frac{4y}{y+1} \frac{1-2yr +r^2}{(1-r)^2} \bigg] |V_{cb}|^2
 {\cal F}(y)^2,
\label{eq:sldiff}
\eea
where $r\equiv M_{D^*}/M_B$ and ${\cal F}(y)$ is an effective semileptonic
form factor. The latter is a calculable function of the $\xi_i(y)$s.
To make contact with experiments, one defines the slope
\be
\hat \rho^2 \equiv - \frac{1}{{\cal F}(1)} \frac{d}{dy} {\cal F}(y) \bigg|_{\, y=1},
\ee
which can be extracted from the measurement of
the semileptonic differential rate.

The relation between the experimental slope $\hat\rho^2$ and the theoretical
parameter $\rho^2$ depends on the model used for the calculation of
eq.~(\ref{eq:sldiff}). We found
\bea
\mbox{LINSR} && \rho^2=\;\hat\rho^2-0.13,\nn\\
\mbox{NRSX} && \rho^2= \;\hat\rho^2+0.084,
\eea
to be compared with the results of ref.~\cite{rhohat},
$\rho^2=\;\hat\rho^2-(0.14\pm 0.02)+{\cal O}(1/m)$ and 
$\rho^2=\;\hat\rho^2 \pm 0.2$ respectively.

In the next section, we will study the dependence of our results on
the physical slope $\hat\rho^2$, rather than $\rho^2$.

\section{Fitting Cabibbo-allowed decay modes}
\label{sec:fit}
%
%
\begin{table}  
\begin{center}
\begin{tabular}{|c|l|r|}
\cline{2-3}
\multicolumn{1}{c|}{~}& Channel & $BR$ $\times 10^5$ \\ 
\hline
				& $\star$  $B_d \to \pi^+ D^{-}$&  $300 \pm 40$	\\
				& $\star$ $B_d \to \pi^+ D^{*-}$&  $276 \pm 21$	\\
				& $\star$ $B_d \to \rho^+ D^{-}$&  $790 \pm 140$	\\
				& $\star$ $B_d \to \rho^+ D^{*-}$& $670 \pm 330$	\\
\cline{2-3}
				& $B_d \to D_s^+ D^{-}$	&  $800 \pm 300$	\\
Type I				& $B_d \to D_s^+ D^{*-}$&  $960 \pm 340$\\
$\propto \vert a_1\vert^2$	& $B^+ \to D_s^+ \bar D^{0}$ & $1300 \pm 400$\\
				& $B^+ \to D_s^+ \bar D^{*0}$& $1200 \pm 500$\\
				& $B_d \to D_s^{*+} D^{-}$& $1000 \pm 500$\\
				& $B_d \to D_s^{*+} D^{*-}$& $2000 \pm 700$\\
				& $B^+ \to D_s^{*+} \bar D^{0}$& $900 \pm 400$\\
				& $B^+ \to D_s^{*+} \bar D^{*0}$&$2700 \pm 1000$\\
\hline
				&  $B_d \to K^0 J/\psi$	&  $89 \pm 12$	\\
Type II 			&  $B_d \to K^{*0} J/\psi$&  $135 \pm 18$	\\
$\propto \vert a_2\vert^2$	&  $B^+ \to K^+ J/\psi$	&  $99 \pm 10$ 	\\
				&  $B^+ \to K^{*+} J/\psi$&  $147 \pm 27$	\\
\hline
				& $\star$ $B^+ \to \pi^+ \bar D^{0}$ & $530 \pm 50$\\
Type III			& $\star$ $B^+ \to \pi^+ \bar D^{*0}$& $460 \pm 40$\\
$\propto \vert x_1 a_1+ x_2 a_2\vert^2$
				& $\star$ $B^+ \to \rho^+ \bar D^{0}$& $1340 \pm 180$\\
				& $\star$ $B^+ \to \rho^+ \bar D^{*0}$& $1550 \pm 310$\\
\hline
\end{tabular}\\[0.2cm]
\begin{tabular}{|lrr|}
\hline 
$q^2$ & $\frac{d}{dq^2} BR(B \to D^* l^+ \nu)(GeV^{-2})$
   & $\frac{d}{dq^2} BR(B \to D l^+ \nu)(GeV^{-2})$ \\
\hline
$m^2_{\pi}$     & $(0.237 \pm 0.026 )\times 10^{-2}$ 	&	$(0.35 \pm 0.06)\times 10^{-2}$ 	\\
$m^2_{\rho}$    & $(0.250 \pm 0.030 )\times 10^{-2}$ 	&	$(0.33 \pm 0.06)\times 10^{-2}$	\\
$m^2_{D_s}$     & $(0.483 \pm 0.033)\times 10^{-2}$ 	&	\multicolumn{1}{c|}{--}	\\
$m^2_{D_s^*}$   & $(0.507 \pm 0.035)\times 10^{-2}$ 	&	\multicolumn{1}{c|}{--}	\\
\hline
\end{tabular}
\label{tab:data}
\caption{{\it Experimental branching ratios \protect\cite{cleofp,pdg,brdq}
of decay modes to be used in
the fit of the parameters $\alpha$, $\xi$ and $\delta_\xi$. The classification
of non-leptonic channels according to their dependence on $a_1$ and $a_2$ is
also shown. The channels marked with $\star$ have been used for the fit
of the ratios defined in eq.~(\protect\ref{eq:slr}).}}
\end{center}
\end{table}

This section contains the results of our phenomenological analysis of
Cabibbo-allowed $B$ decays, focused on the r\^ole of heavy-heavy
form factors. We proceed as follows:
\begin{itemize}
\item we show that the best fit to the $BR(B\to D\pi)$ and $BR(B\to KJ/\Psi)$   is
obtained for different values of $\hat\rho^2$, depending on the
model used for computing heavy-heavy form factors;
\item we use the ratios $R_\pi(B\to D\pi)$ introduced in
ref.~\cite{ratio}, see below,  and
show that, by using them instead of the $BR(B\to D\pi)$, it is possible
to fit the FP (almost) independently of $\hat\rho^2$. This method gives
our best determination of the FP, free from theoretical uncertainties coming
from the assumptions made for the heavy-heavy form factors;
\item using the ($\hat\rho^2$-independent) FP, we perform a fit to the 
$BR(B\to D\pi)$ in order to extract a preferred range for the value of
 $\hat\rho^2$; the results  are model
dependent, in agreement with our first finding;
\item we show that the different ranges of $\hat\rho^2$ actually correspond
to the same values of the relevant  form factors at $q^2=M_\pi^2\sim
0$. Using the HQET, the latters can be determined by factorization applied to
$B\to D\pi$ decays and may be compared with direct measurements 
from semileptonic decays.
\end{itemize}

The relevant decay modes which we use in the fits are listed in
tab.~\ref{tab:data}. It is well known that, in the factorization approximation,
only two combinations of $DE$ and $CE$ appear in the amplitudes of these
decays.
This feature is taken into account by the parameters $a_1$ and
$a_2$ introduced by BSW~\cite{bsw87}. The relation between $a_1$ and $a_2$
and our parameters is given by
\be
a_1=\alpha \left(C_2+\xi e^{i\delta_\xi} C_1\right),\qquad
a_2=\alpha \left(C_1+\xi e^{i\delta_\xi} C_2\right)
\label{eq:a1a2}
\ee 
where $C_1$ and $C_2$ are the Wilson coefficients of the operators defined
in eq.~(\ref{eq:oper})~\footnote{Notice that our operator basis
differs from the one of ref.~\cite{bsw87} by the trivial exchange
$Q_{1,2}\leftrightarrow Q_{2,1}$.}.
In tab.~\ref{tab:data}, the decay
modes are organized according to the standard classification in three
classes. Amplitudes of Type I, II and III decay modes are proportional to
$\vert a_1\vert$, $\vert a_2\vert$ and $\vert x_1 a_1+x_2 a_2\vert$
respectively, where $x_i$ are generic, process-dependent coefficients.

Given the structure of the amplitudes, we have to fit decay modes of
all the three classes in order to fully determine the FP.
Note that the three classes have a different
dependence on the form-factors.
While Type-II decays always involve a heavy-light transition,
heavy-heavy form factors enter Type-I modes only. The latter  is a
general feature, since Type-I transitions are always driven by
charged currents and are therefore proportional to $a_1$. In general Type-III
modes involve transitions of both sorts. In our parameterization,
Type-I modes essentially fix $\alpha$, while both Type II and III are needed
to constrain $\xi$ and $\delta_\xi$. 

\begin{figure}[t]
\centering
\begin{minipage}  {0.05\linewidth}
\vspace*{-10.0cm}
\large{$\frac{\chi^2}{dof}$}
\end{minipage}
\epsfig{file=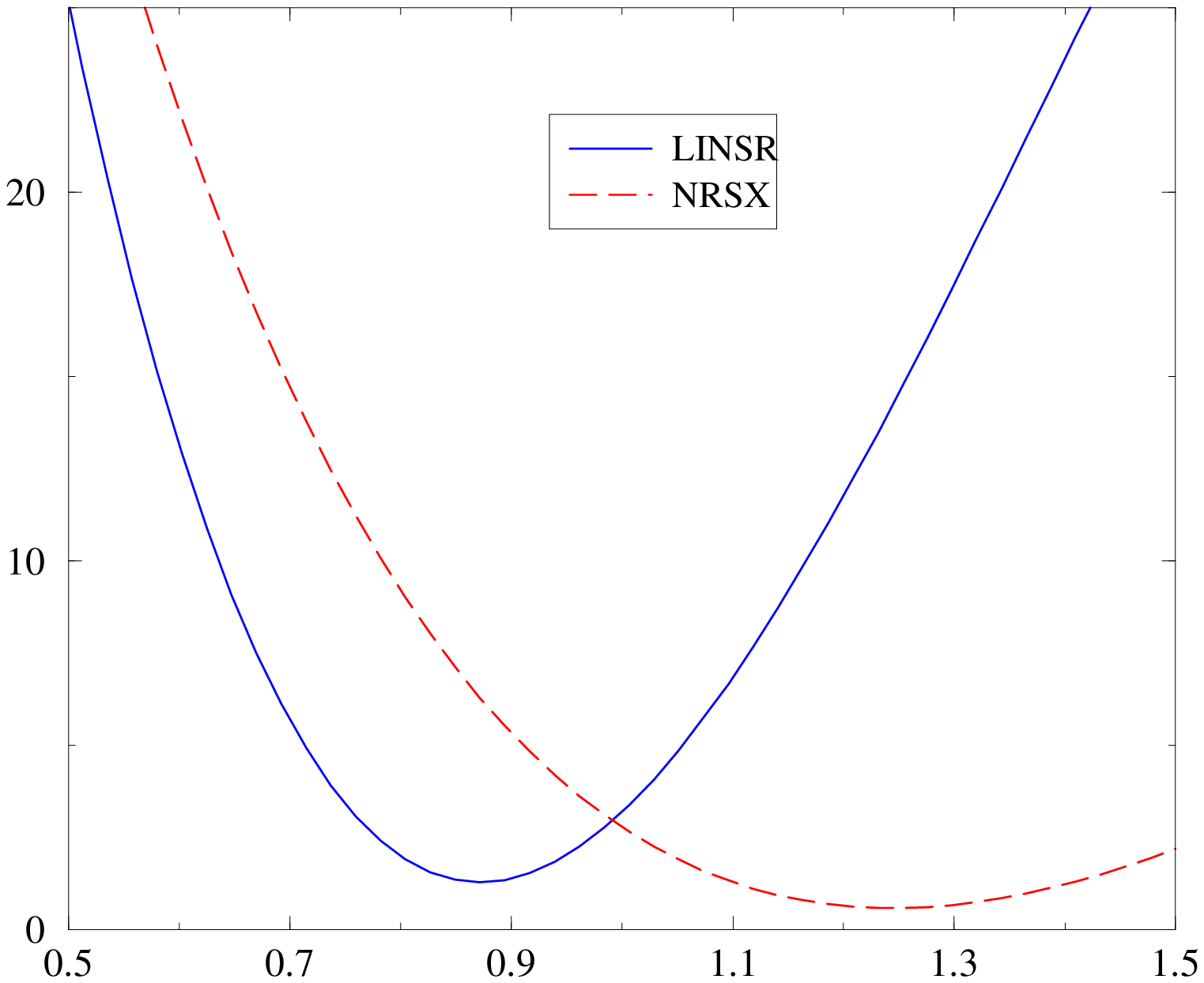,height=0.4\textheight} \\[5pt]
\hspace*{1.6cm}$\hat\rho^2$
\caption{{\it $\hat\rho^2$ dependence of the minimum $\chi^2/dof$ from the fit of $\xi$
to $BR(B\to D\pi)$ and $BR(B\to KJ/\Psi)$ assuming $\alpha=1$
and $\delta_\xi=0$. Both LINSR and NRSX results
are shown.}}
\label{fig:rhodep}
\end{figure}

Contrary to the common wisdom, the assumptions made
on the momentum dependence of the  heavy-heavy form factors introduce large 
uncertainties in the determination of the FP from this fit. 
This is more easily shown fitting only the effective number of colour as done
in the old literature \footnote{Had we fitted also $\alpha$, the resulting
minimum $\chi^2/dof$ would have been almost independent of $\hat\rho^2$,
since $\alpha$ easily compensates the variation of the form factors with $\hat\rho^2$.
Still, the fitted value of $\alpha$ would have been strongly $\hat\rho^2$ dependent.}.
In our parametrization this corresponds to assume
$\alpha=1$ and $\delta_\xi=0$ and to fit $\xi$ only. 
The result of this fit is shown in fig.~\ref{fig:rhodep}, where minimum values of
$\chi^2/dof$ from the fit of $\xi$ to the $BR(B\to D\pi)$ and to the $BR(B\to KJ/\Psi)$ 
are plotted as function of $\hat\rho^2$. It is apparent that the best fit is
obtained for quite different values of $\hat\rho^2$, and corresponds to different
values of $\xi$, depending on the model used
to compute the heavy-heavy form factors. Consequently, in general, the FP fitted
using $BR$s at fixed $\hat\rho^2$, as usually done in the literature, suffer from
a large theoretical error, which was previously hidden in the choice of 
a specific
model when fitting the data. The second important remark is that a comparison of
the value of $\hat\rho^2$,  the ``physical" slope
measured in semileptonic decays, with that  extracted from non-leptonic decays
is not a good test of factorization  since, in the latter case, the result is 
model dependent.
\par To circumvent this problem in the determination of  the FP,
instead of the Type-I and Type-III $BR$s,   we fit the ratios
\be
R_M(B\to D^{(*)}M)=\frac{BR(B\to D^{(*)} M)}{\frac{d}{dq^2}BR(B\to D^{(*)}
l\nu)\vert_{q^2=m_M^2}},
\label{eq:slr}
\ee
where $M=\pi,D_s,\dots$ is the emitted meson. 
The advantage of using eq.~(\ref{eq:slr}) is that in these ratios the heavy-heavy
form factor dependence drops out completely for Type-I decays and is strongly
reduced for Type III. In practice,  we used  the  ratios
corresponding to the non-leptonic decays marked with $\star$ in 
table~\ref{tab:data}. In the fit, besides  the ratios $R_M$ for Type-I and Type-III,
we also use all the $BR$s of  the Type-II decays. 
\begin{table}[t]
\begin{center}
\begin{tabular}{c|c|r|r|r|}
\hline
\multicolumn{2}{|c|}{LINSR} & $\hat \rho^2=0.80$ 
	& $\hat \rho^2=0.90$ & $\hat \rho^2=1.00$ \\
\hline
$R_{\pi}(B\to D\pi)$ & $\chi^2/dof$ & $1.36$ & $1.36$ & $1.36$  \\
Type I & $\alpha$ & $1.02$ & $1.02$ & $1.02$ \\
\multicolumn{1}{c|}{+} & $\xi$ & $0.45$ & $0.45$ & $0.45$ \\
$BR(B\to K J/\psi)$ & $\delta_\xi$ & $0.00$ & $0.00$ & $0.00$ \\
\hline
$R_{\pi}(B\to D\pi)$ & $\chi^2/dof$ & $1.40$ & $1.39$ & $1.42$ \\
Type I+III & $\alpha$ & $1.05$ & $1.04$ & $1.03$\\
\multicolumn{1}{c|}{+} & $\xi$ & $0.44$ & $0.44$ & $0.44$ \\
$BR(B\to K J/\psi)$ & $\delta_\xi$ & $0.00$ & $0.00$ & $-0.26$ \\
\cline{2-5}
\end{tabular}\\[.2cm]
\begin{tabular}{c|c|r|r|r|r|r|}
\hline
\multicolumn{2}{|c|}{NRSX} & $\hat \rho^2=1.25$
	& $\hat \rho^2=1.35$ & $\hat \rho^2=1.45$ \\
\hline
$R_{\pi}(B\to D\pi)$ & $\chi^2/dof$ & $0.39$ & $0.39$ & $0.39$  \\
Type I & $\alpha$ & $1.01$ & $1.01$ & $1.01$ \\
\multicolumn{1}{c|}{+} & $\xi$ & $0.38$ & $0.38$ & $0.38$ \\
$BR(B\to K J/\psi)$ & $\delta_\xi$ & $0.00$ & $0.00$ & $0.00$ \\
\hline
$R_{\pi}(B\to D\pi)$ & $\chi^2/dof$ & $0.71$ & $0.70$ & $0.68$ \\
Type I+III & $\alpha$ & $1.04$ & $1.04$ & $1.03$ \\
\multicolumn{1}{c|}{+} & $\xi$ & $0.38$ & $0.38$ & $0.38$ \\
$BR(B\to K J/\psi)$ & $\delta_\xi$ & $0.00$ & $0.00$ & $0.00$ \\
\cline{2-5}
\end{tabular}
\end{center}
\caption{{\it Results of the fit using $D\pi$ semileptonic ratios and
$BR(B\to KJ/\Psi)$ for three different values of the slope
$\hat\rho^2$ and the two form-factor models, LINSR and NRSX, described in the
text. For each model, two fits have been performed, the difference being
the inclusion of Type-III $D\pi$ modes, which introduce some $\hat\rho^2$
dependence.}}
\label{tab:fit}
\end{table}

The results of the fit are shown in tab.~\ref{tab:fit} for NRSX and LINSR.
We do not include the $D^{(*)}D_s^{(*)}$ modes for two reasons: on the one hand, 
their contribution to the total $\chi^2$ is suppressed by the large 
experimental errors
in the measured $BR$s;
on the other, we want to use them to extract the decay constants $f_{D_s}$ and $f_{D_s^*}$.

For both choices of form factors, we give the results of two different fits:
the first includes all types of decays and determines the FP
$\alpha$, $\xi$ and $\delta_\xi$. It retains, however, a small residual dependence on
heavy-heavy form factors, {\it i.e.} on $\hat \rho^2$. 
The second is a fit to Type-I and -II channels only, which is 
totally independent of $\hat\rho^2$. The results are quite close.
Note that the second fit only involves  two combinations of the three FP. 
As a consequence we have to fix one parameter in order to extract the other two:
we choose to put $\delta_\xi=0$, quite consistently with what has been found 
with the first fit. As a consistency check, we have also verified that different values
of $\delta_\xi$ do not change appreciably the results~\footnote{
The fit is not very sensitive to $\delta_\xi$, thus it cannot fix this
parameter very precisely, see the final results in~(\ref{eq:fitres}).}.
In tab.~\ref{tab:fit}  we show the fitted values of  the FP 
for several choices  of $\hat\rho^2$,  for both NRSX and LINSR.  As mentioned
above, the results   turn out to be,  within a given model, independent of $\hat\rho^2$.
\begin{figure}[t]  
\centering
\begin{minipage}  {0.05\linewidth}
\vspace*{-10.0cm}
$\hat \rho^2$
\end{minipage}
\epsfig{file=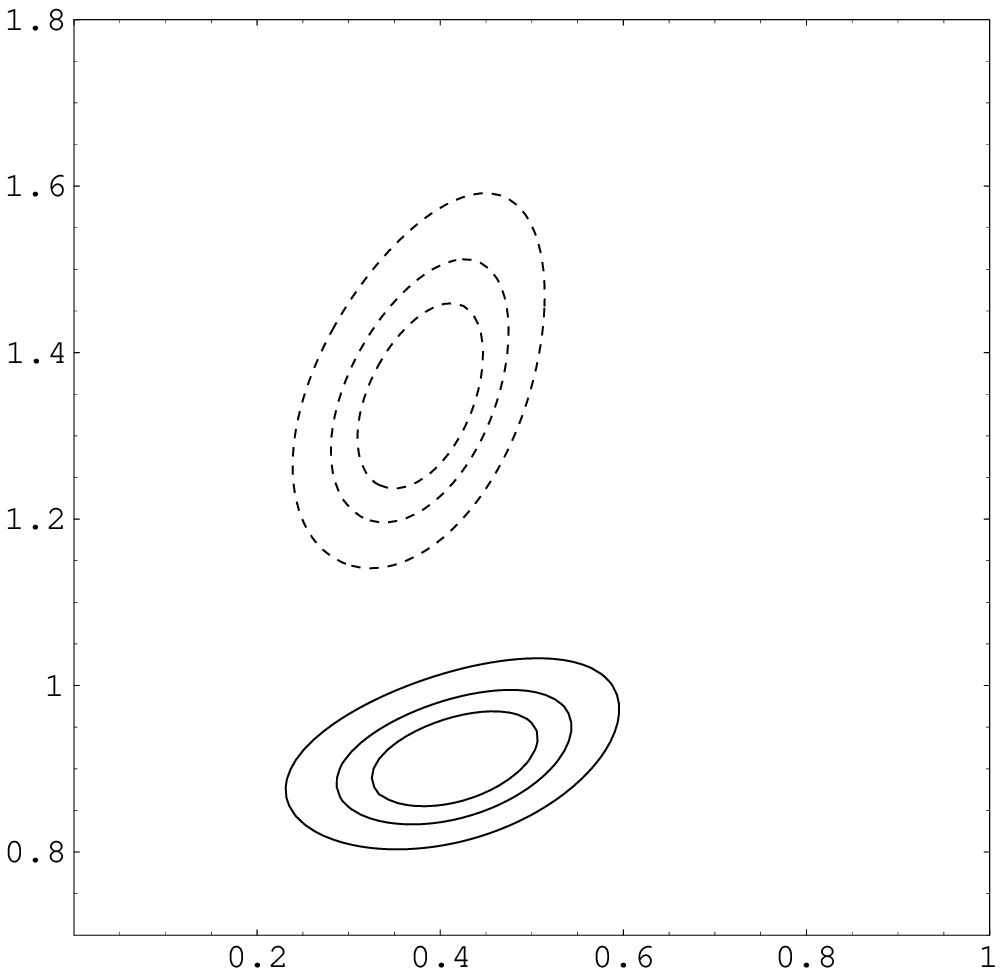,width=0.6\linewidth} \\[5pt]
\hspace*{1.6cm}$\xi$
\caption{{\it Contour plots of $\chi^2/dof$ in the plane $(\xi,
\hat\rho^2)$, obtained from a two-parameter fit to the
non-leptonic $B\to D\pi$ decays. The other parameters ($\alpha$ and
$\delta_\xi$) have been fixed by the fit of tab.~\protect\ref{tab:fit} in
a $\hat\rho^2$-independent way. The solid curves refer to LINSR,
the dashed ones to NRSX. The different contours correspond to
$\chi^2/dof=(1.5,2,3)\times\chi^2_{min}/dof$.}}
\label{fig:rho2xi}
\end{figure}

Having fitted the FP in a $\hat\rho^2$-independent way, we now use
the $BR(B\to D\pi)$ to extract from the data a preferred range of
$\hat\rho^2$. Notice that $\delta_\xi$ is not a critical parameter, since
the results of the fits are not very sensitive to its value,
and that the values of  $\alpha$ and $\hat\rho^2$ are trivially correlated,
because the amplitudes only depend on the product of $\alpha$ with the
effective form factors at $q^2 \sim 0$. Therefore we choose to perform
a two-parameter fit of $\xi$ and $\hat\rho^2$ using the $BR(B\to D\pi)$, at
fixed  values of $\delta_\xi$ and $\alpha$, as extracted from the previous
fit of tab.~\ref{tab:fit}. In this way, we can study the correlations in the
$(\xi$--$\hat\rho^2)$ plane and check the consistency of the determination of
$\xi$ using different fitting procedures.

Figure~\ref{fig:rho2xi} shows the contour plots of $\chi^2/dof$ in
the $(\xi,\hat\rho^2)$ plane for NRSX and LINSR. The fitted value of
$\xi$ is consistent with tab.~\ref{tab:fit} and the preferred $\hat\rho^2$
is larger using NRSX than LINSR. Moreover, the LINSR $BR$s are steeper
functions of $\hat\rho^2$, consistently with fig.~\ref{fig:rhodep}.
This observation justifies the choice of the set of values  of $\hat\rho^2$ used in
tab.~\ref{tab:fit}. 

\begin{figure}[t]
\begin{minipage}{0.43\linewidth} 
{\small $f_+(M_\pi^2)$} \\[-1cm]  \hspace*{1.2cm} 
\epsfig{file=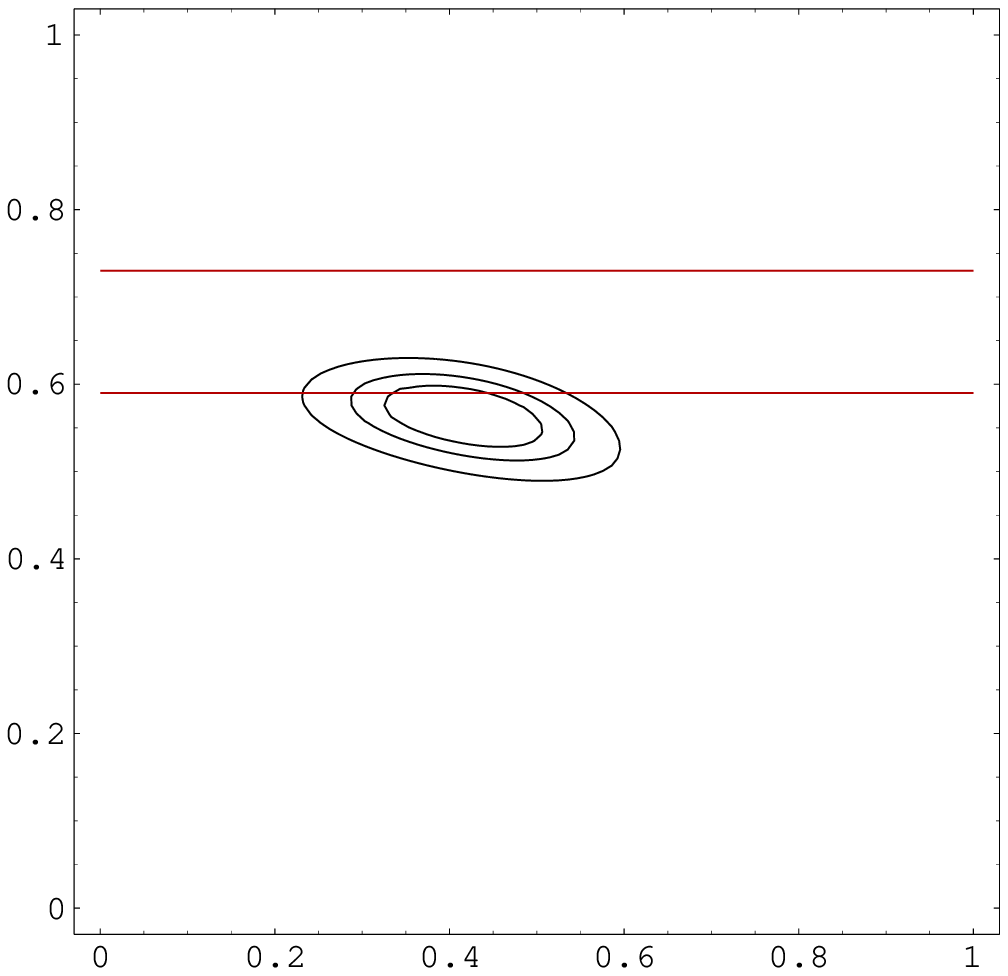,width=0.8\linewidth} \\
\hspace*{4cm} {\small $\xi$} 
\end{minipage} \hspace*{1.0cm}
\begin{minipage}{0.43\linewidth} 
{\small $f_+(M_\pi^2)$} \\[-1cm]  \hspace*{1.2cm} 
\epsfig{file=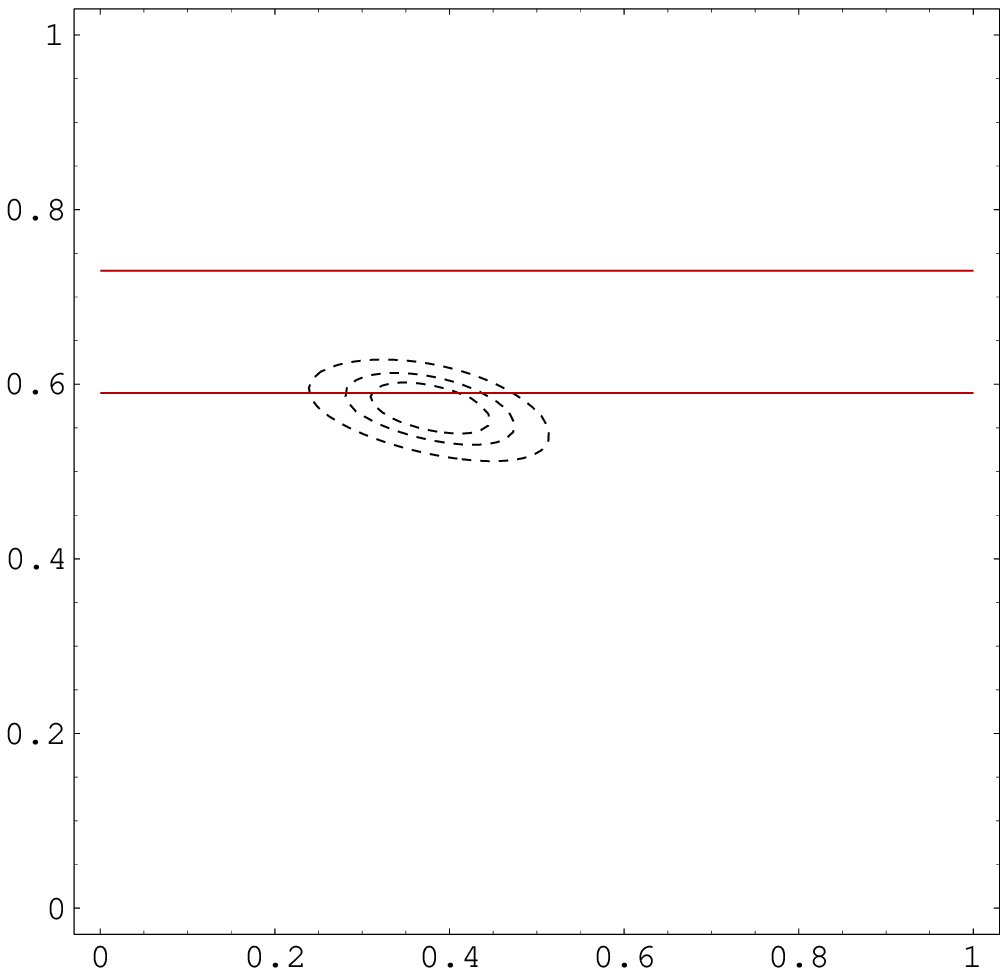,width=0.8\linewidth} \\
\hspace*{4cm} {\small $\xi$} 
\end{minipage} \\[1cm]
\begin{minipage}{0.43\linewidth} 
{\small $V(M_\pi^2)$} \\[-1cm]  \hspace*{1.2cm} 
\epsfig{file=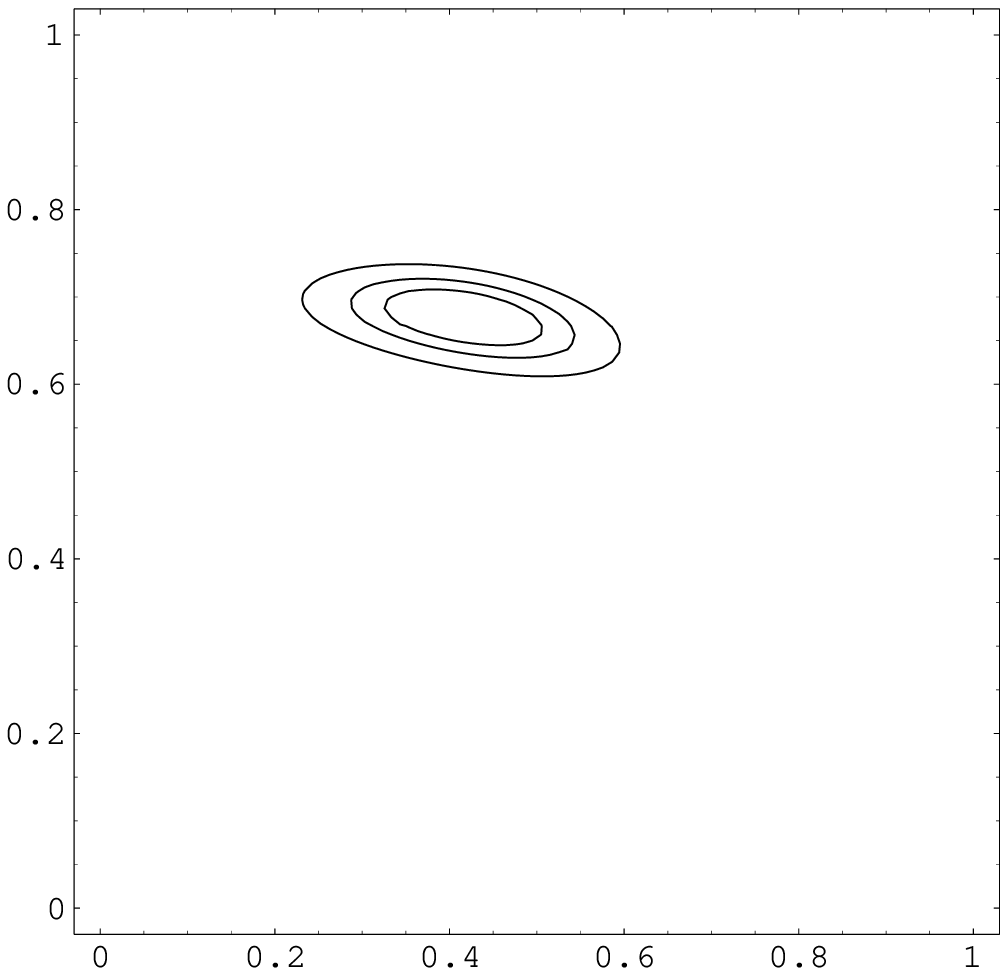,width=0.8\linewidth} \\
\hspace*{4cm} {\small $\xi$} 
\end{minipage} \hspace*{1.0cm}
\begin{minipage}{0.43\linewidth} 
{\small $V(M_\pi^2)$} \\[-1cm]  \hspace*{1.2cm} 
\epsfig{file=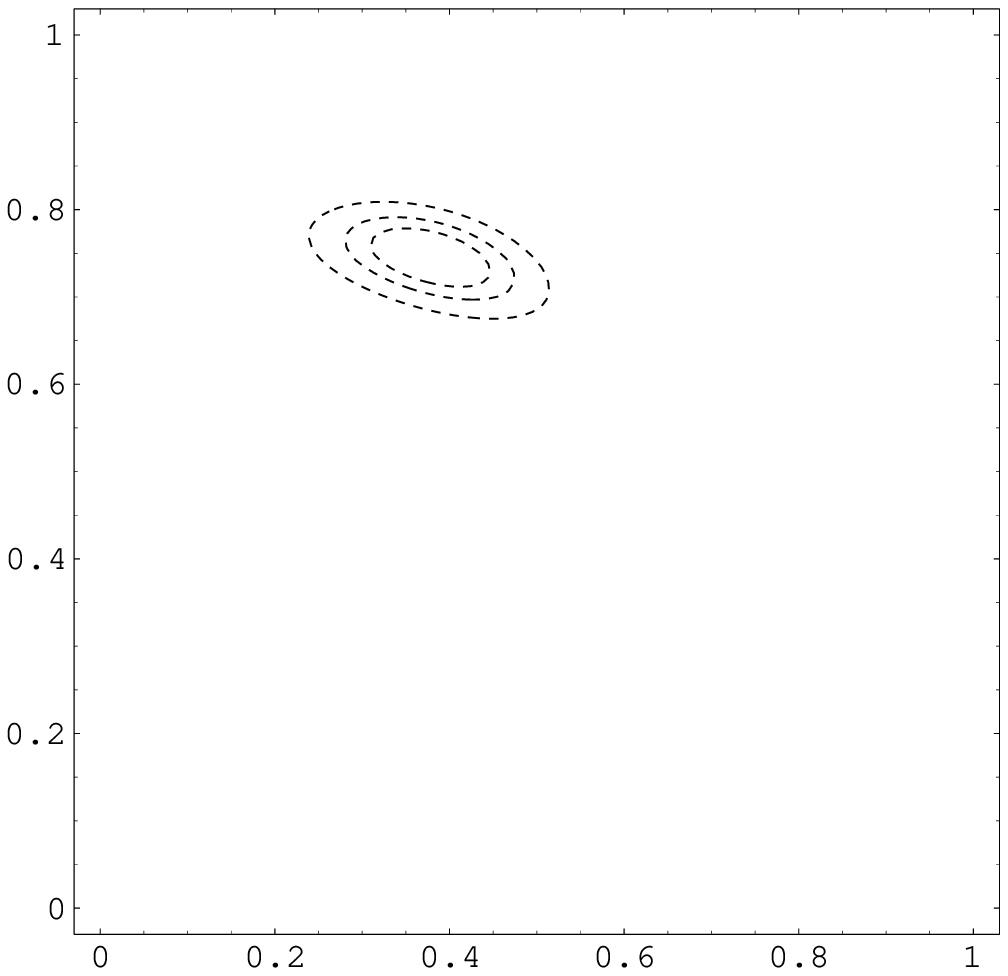,width=0.8\linewidth} \\
\hspace*{4cm} {\small $\xi$} 
\end{minipage} \\[1cm]
\caption{{\it
Determination of various $B\to D\pi$ form factors at $q^2=M_\pi^2$
as functions of $\xi$. These plots
are obtained from fig.~\protect\ref{fig:rho2xi} using the relations connecting
heavy-heavy form factors to the Isgur-Wise function,
eqs.~(\protect\ref{eq:Avett})--(\protect\ref{eq:hhff}). LINSR (NRSX) form
factors are shown in the left (right) column.
The different contours correspond to
$\chi^2/dof=(1.5,2,3)\times\chi^2_{min}/dof$.}}
\label{fig:ff1}
\end{figure}

\begin{figure}
\begin{minipage}{0.43\linewidth} 
{\small $A_0(M_\pi^2)$} \\[-1cm]  \hspace*{1.2cm} 
\epsfig{file=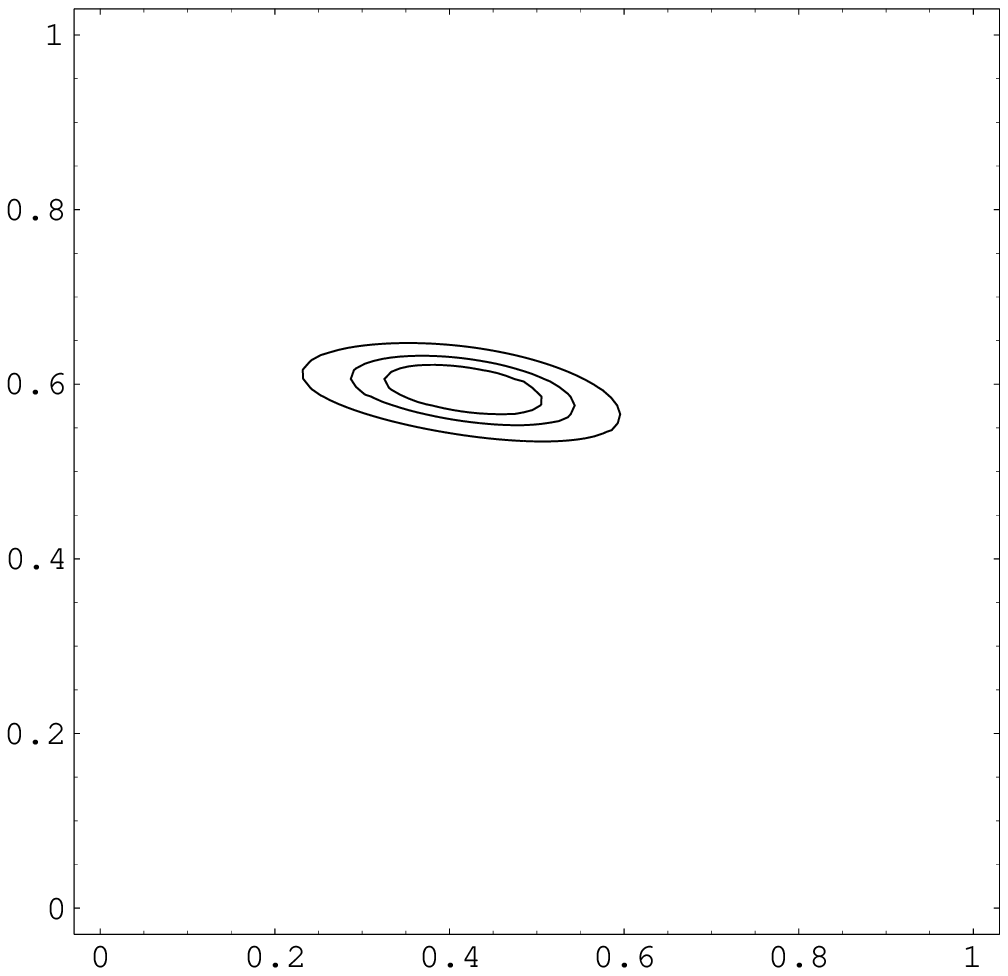,width=0.8\linewidth} \\
\hspace*{4cm} {\small $\xi$} 
\end{minipage} \hspace*{1.0cm}
\begin{minipage}{0.43\linewidth} 
{\small $A_0(M_\pi^2)$} \\[-1cm]  \hspace*{1.2cm} 
\epsfig{file=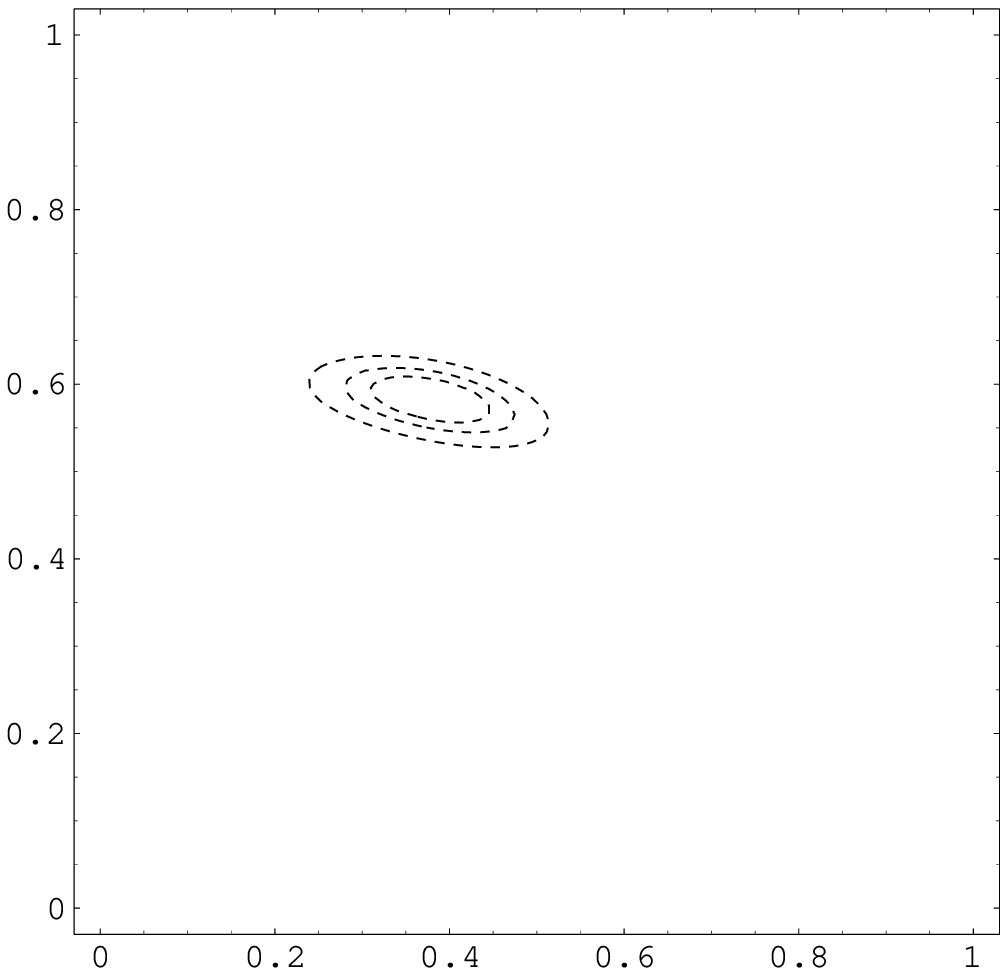,width=0.8\linewidth} \\
\hspace*{4cm} {\small $\xi$} 
\end{minipage} \\[1cm]
\begin{minipage}{0.43\linewidth} 
{\small $A_1(M_\pi^2)$} \\[-1cm]  \hspace*{1.2cm} 
\epsfig{file=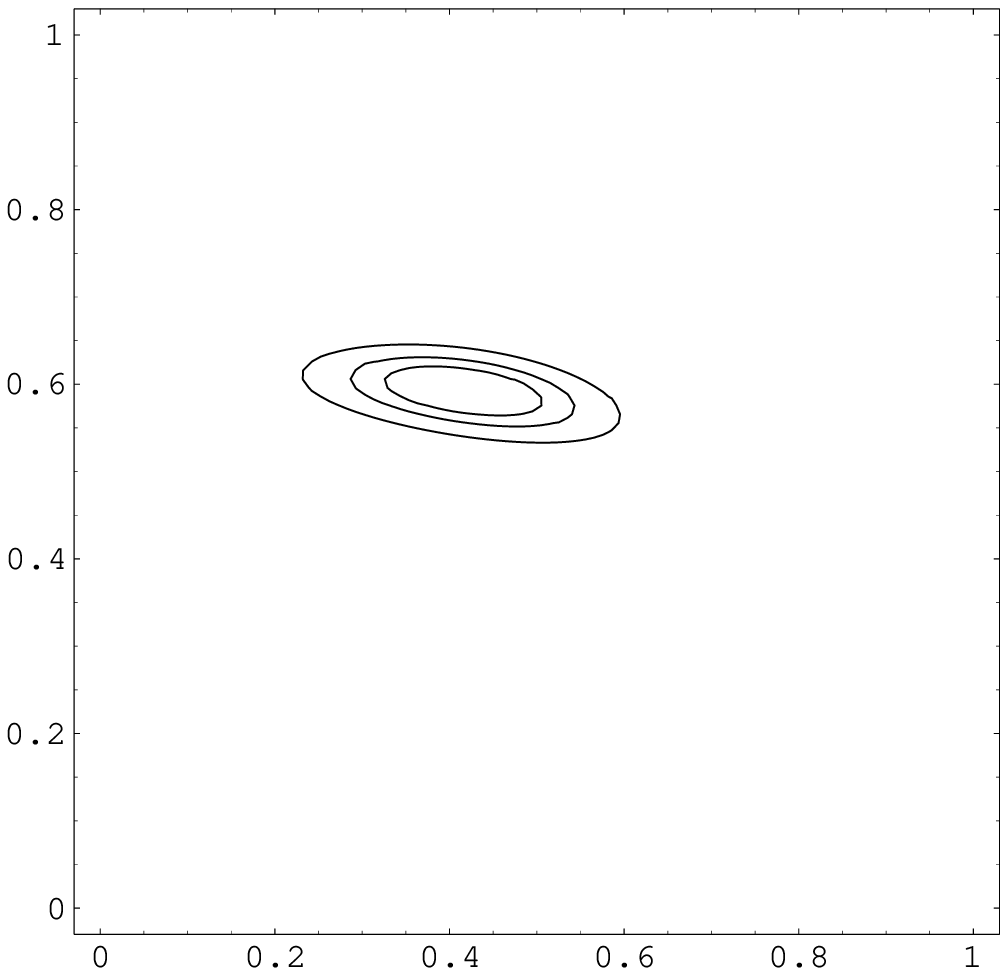,width=0.8\linewidth} \\
\hspace*{4cm} {\small $\xi$} 
\end{minipage} \hspace*{1.0cm}
\begin{minipage}{0.43\linewidth} 
{\small $A_1(M_\pi^2)$} \\[-1cm]  \hspace*{1.2cm} 
\epsfig{file=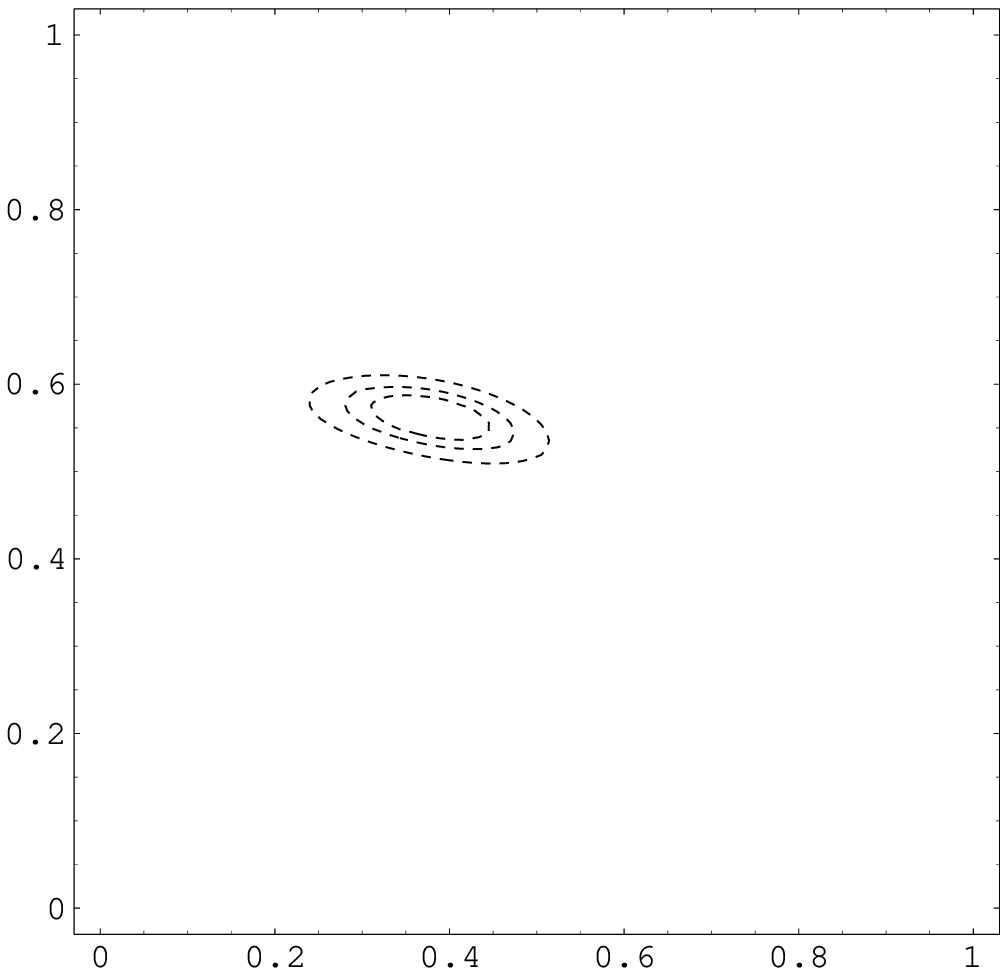,width=0.8\linewidth} \\
\hspace*{4cm} {\small $\xi$} 
\end{minipage} \\[1cm]
\begin{minipage}{0.43\linewidth} 
{\small $A_2(M_\pi^2)$} \\[-1cm]  \hspace*{1.2cm} 
\epsfig{file=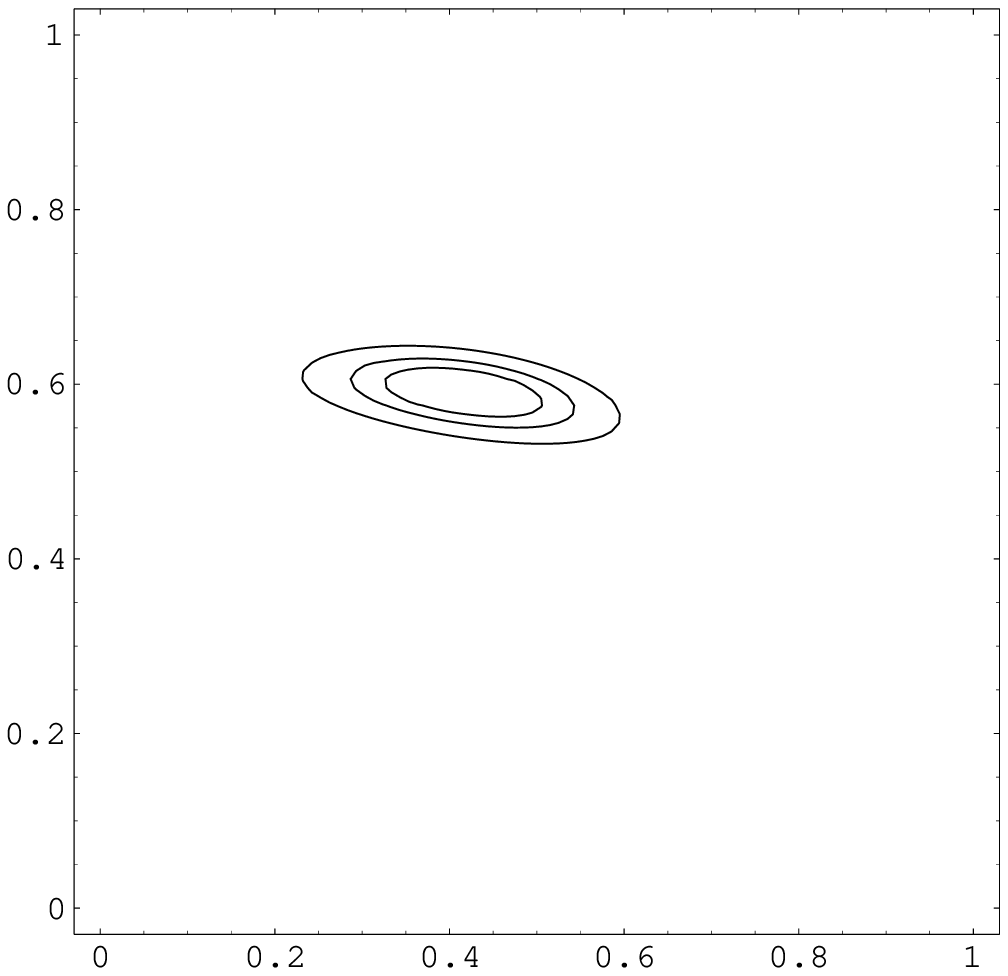,width=0.8\linewidth} \\
\hspace*{4cm} {\small $\xi$} 
\end{minipage} \hspace*{1.0cm}
\begin{minipage}{0.43\linewidth} 
{\small $A_2(M_\pi^2)$} \\[-1cm]  \hspace*{1.2cm} 
\epsfig{file=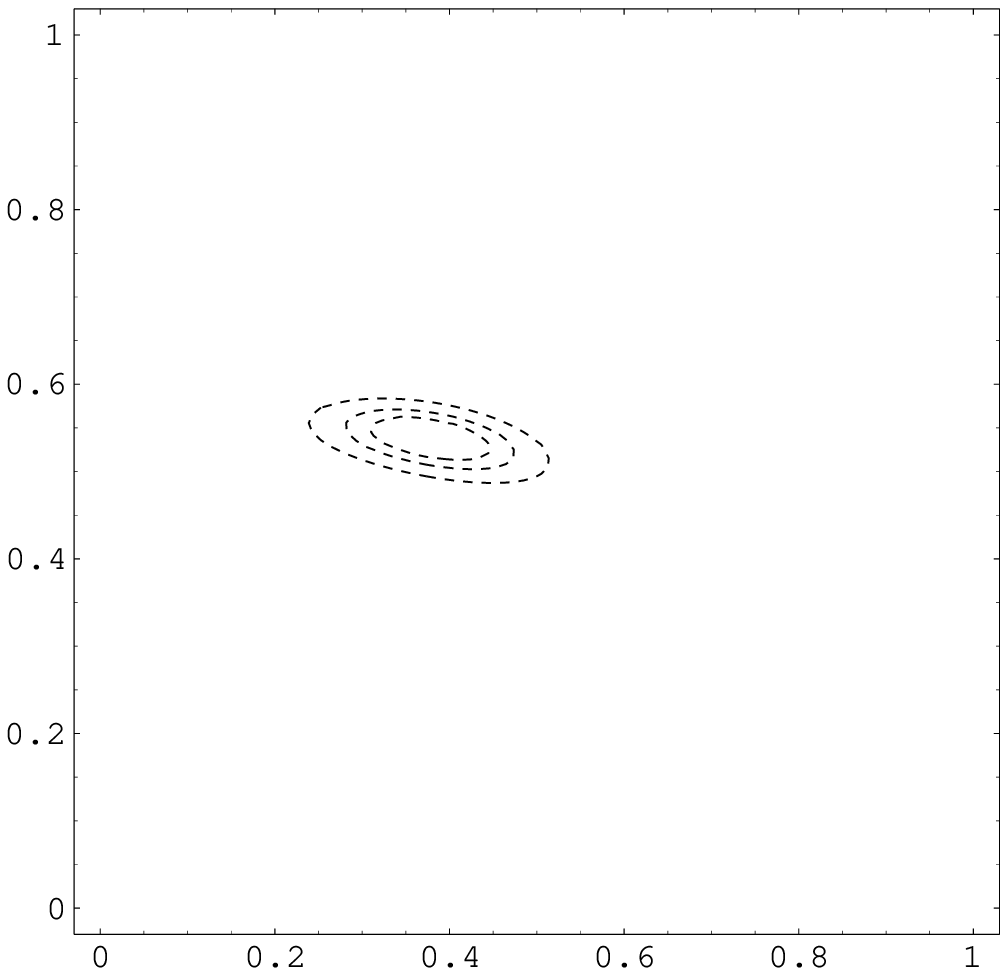,width=0.8\linewidth} \\
\hspace*{4cm} {\small $\xi$} 
\end{minipage}

\caption{{\it The same plots as in fig.~\protect\ref{fig:ff1} for 
$A_0$, $A_1$ and $A_2$.}}
\label{fig:ff2}
\end{figure}

It is not surprising that the  fit to the $BR(B\to D\pi)$ gives  values of $\hat\rho^2$
which are   model dependent. The fit only  fixes the values of the relevant
heavy-heavy form factors $f_i=f_+,A_0,\dots$ at
$q^2=M_\pi^2\sim 0$~\footnote{Type-III modes actually depend also
on heavy-light form factors, which however appear in color suppressed
contributions to the total amplitude.}. These form factors can be expressed
in terms of the $\xi_i$s at $q^2 \sim 0$ which, in turn, are related to
the IW function  $\xi(y)$
by heavy quark symmetry, see eq.~(\ref{eq:hqetff}).
The relation between the fitted form factors at $q^2=M_\pi^2$ and the values of
$\xi_i(1)$, which are fixed by the HQET,
depends on  the functional form adopted for the IW function  $\xi(y)$ and  
on  the value of  $\rho^2$.
Thus different values of $\hat\rho^2$ are obtained  
by fitting the data with different models. In particular,
we find that  the main difference
between NRSX and LINSR relies on the choice of the IW function,
eqs.~(\ref{eq:LINSRIW}) and (\ref{eq:NRSXIW}), rather than in the inclusion of
$1/m$ corrections.
Plotting the results of the fit   in
the planes $(\xi,f_i)$, through eqs.~(\ref{eq:Avett})--(\ref{eq:hhff}),
we obtain almost the same determination of the $f_i(M_\pi^2)$ 
with both NRSX and LINSR, as shown in figs.~\ref{fig:ff1} and \ref{fig:ff2}.
Although we have considered only two models in the present study, we believe that 
this result is quite general. \par 

We stress again that constraints on the heavy-heavy form
factors  can only be obtained by combining the results
of two independent fits: the first which fixes the FP using   the ratios $R_\pi$, that
are essentially independent on the model used to calculate the form
factors, and  the second which fixes the form factors using the $BR(B\to D\pi)$
at fixed values
of the FP.

The comparison of the heavy-heavy form factors directly measured in semileptonic decays
at $q^2=M_\pi^2\sim 0$ with
the results in figs.~\ref{fig:ff1} and \ref{fig:ff2}, and table~\ref{tab:ff},
is a real test of generalized factorization in $B\to D\pi$
decays, independently on the choice of the IW function and of
the value of $\hat\rho^2$. This checks the assumptions we made for
computing hadronic matrix elements as described in the previous
sections. Since we use the HQET relations coming from
eq.~(\ref{eq:hqetff}), we are left with only one independent form factor, namely
the IW function. Therefore the only form factor directly measured at $q^2=0$,
$f_+(0)$, already allows a test of our approach. Its value,
$f_+(0)=0.66\pm 0.06\pm 0.04$~\cite{cleofp}, showed as a band in the upper
plots of fig.~\ref{fig:ff1}, agrees well with the result of the fit.
The extraction of the other form factors at $q^2=0$
from the CLEO data is under way~\cite{artuso} and would allow a check of
the HQET relations among $B\to D$ form factors near the maximum recoil
point. Notice that, at least in principle, the fitting procedure described in
this section could be used to extract independently
the values of all the form factors
at $q^2=M_\pi^2$, just including them among the FP. In this case, the
comparison of each form factor with the measurements from semileptonic decays
would be a test of generalized factorization, independent of the HQET
relations eq.~(\ref{eq:hqetff}).
Unfortunately the present accuracy of the data does not allow a
separate determination of the different form factors.

From the discussion above, we conclude that the HQET-inspired parameterization
of the heavy-heavy form factors in terms of their  value at $q^2_{max}$ and of
the slope $\hat\rho^2$, which is commonly adopted by experimental collaborations
and successfully applied to semileptonic decays, is not the most appropriate choice
for the factorization analysis of non-leptonic decays.

With the results for the FP given in  (\ref{eq:fitres}), we can predict $BR$s
of yet-unmeasured decay channels, having one $D$ and one light meson
in the final state. We list our  predictions in tab.~\ref{tab:pred},
where the ranges in square brackets give an estimate of the
theoretical uncertainties. They  were found by allowing 
values of $\chi^2/dof$ up to three times larger than the minimum.
Flavor $SU(3)$ symmetry justifies the use of parameters
obtained from $B\to D\pi$ and $B\to KJ/\Psi$ decays to the decays listed in table~\ref{tab:pred}.
Large flavour effects are unlikely, since the factorized amplitudes already account
for some $SU(3)$ breaking.

Finally, we summarize the result of our fits of the FP by quoting their  best values and
ranges of variation, obtained  by allowing values of $\chi^2/dof$
up to three times larger than the minimum. The comparison
between the two different models, NRSX and LINSR, gives us an estimate
of the theoretical uncertainties due to the form-factor model dependence. 
As discussed before, these  FP parameters are those obtained using the coefficients functions
computed at the NLO in the $\overline{\rm MS}$ scheme with $\mu=5$ GeV. 
We obtain
\be
\begin{tabular}{|c|rr|rr|}
\cline{2-5}
\multicolumn{1}{c|}{} & \multicolumn{2}{c|}{LINSR} & \multicolumn{2}{c|}{NRSX}\\
\hline
$\alpha$ & 1.04 & [0.9--1.2] & 1.04 & [1.0--1.1]\\
$\xi$ & 0.44 & [0.2--0.5] & 0.38 & [0.2--0.4]\\
$\delta_\xi$ & 0.0 & [-1.5--1.5] & 0.00 & [-1.0--1.0]\\
\hline
$\vert a_1\vert$ & 1.04 & [0.9--1.3] & 1.05 & [1.0--1.2]\\ 
$\vert a_2\vert$ & 0.31 & [0.0--0.7] & 0.25 & [0.0--0.4]\\
\hline
$\frac{\chi^2}{dof}$ & 1.4 & [1.4--4.2] & 0.7 & [0.7--2.1]\\
\hline
$\hat \rho^2$ & 0.91 & [0.8--1.0] & 1.34 & [1.1--1.5]\\
\hline
\end{tabular}
\label{eq:fitres}
\ee
For the sake of comparison with previous literature, we have also
shown the values of $\vert a_1\vert$ and $\vert a_2\vert$, computed using
eq.~(\ref{eq:a1a2}).
It is worth noticing that exact factorization, namely $\alpha=1$,
$\xi=1/3$ and $\delta_\xi=0$, would give  values of $\chi^2/dof$
 $3$--$4$ times larger than the fits which use the  generalized factorization,
 for both the models considered here.

\section{Decay constants from $DD_s$ decays}
\label{sec:costdec}
In this section, we extract the meson decay constants $f_{D_s}$ and
$f_{D_s^*}$ and compare the results with available measurements and
lattice results.

We consider the semi-leptonic ratios $R_{D_s}(B\to D^{(*)}D_s^+)$ and
$R_{D_s^*}(B\to D^{(*)}D_s^{*+})$, introduced in the previous section,
and define the non-leptonic ratios~\cite{neuste}
\bea
R^{P}_{D_s}(B\to D^{(*)}D_s^+)&=&\frac{BR(B\to D^{(*)}D_s^+)}
{BR(B\to D^{(*)}\pi^+)},\nonumber\\
R^{V}_{D_s^*}(B\to D^{(*)}D_s^{*+})&=&\frac{BR(B\to D^{(*)}D_s^{*+})}
{BR(B\to D^{(*)}\rho^+)}.
\label{eq:nlr}
\eea
Up to color-suppressed terms, the factorized amplitudes of
the decay modes considered here are  proportional to $f_{D_s}/f_\pi$ or to
$f_{D^*_s}/f_\rho$. 
Whereas the ratios $R_M$ of eq.~(\ref{eq:slr}) are defined in such a way that the
main form-factor dependence drops out, in the non-leptonic
ratios of eqs.~(\ref{eq:nlr}), the form factors
appearing in the numerator and denominator are evaluated
at different $q^2$ and do not cancel out.
In this case, however, it is the dependence on the FP that
tends to cancel, as long as penguin contributions are neglected.
The  non-leptonic ratios above are exactly independent of $\xi$ and $\delta_\xi$
only if charged $B^+$ decays are not considered, as done in ref.~\cite{neuste}.
In our case, we prefer to double the number of channels in the fit,
by including  charged $B^+$ decays, at the cost of introducing a small
dependence on FP and on the $D^{(*)}$ decay constants, both appearing
in color suppressed terms in the decay amplitudes. We take $f_D=200$ MeV
and $f_{D^*}=220$ MeV~\cite{fdS}.
\par In general, all $DD_s$ modes suffer from a further theoretical error.
This uncertainty originates from using the same FP,
obtained from the fit of sec.~\ref{sec:fit},
in the calculation of the relevant $BR$s
entering the non-leptonic ratios. Since in $DD_s$ decays, the emitted meson is
heavy, one may expect, according to the LEET approach, larger violations to
the factorization limit. In other words, in this case the FP
may significantly differ from those fixed by the $D\pi$ and $K J/\Psi $ modes.
This is a further source of theoretical uncertainty, which we are not able
to estimate at present.

Using the  the semileptonic ratios $R_{D_s}(B\to D^{(*)}D_s^+)$ and
$R_{D_s^*}(B\to D^{(*)}D_s^{*+})$ and the non-leptonic ratios  of eqs.~(\ref{eq:nlr}), 
the form factors
determined from the fit to the $BR(B \to D \pi)$  
and the FP from~(\ref{eq:fitres}),
we have extracted $f_{D_s}$ and
$f_{D_s^*}$. Results are collected in tab.~\ref{tab:costdec}, where
we have separately shown the uncertainties coming from the experimental
errors on the $BR$s and from the errors  on the FP.
As before, in order to
estimate this source of theoretical uncertainty,
we present results obtained using both LINSR and NRSX. 

From tab.~\ref{tab:costdec}, we quote
\be
f_{D_s}=270\pm 35~\mbox{MeV},\qquad f_{D_s^*}=260\pm 30~\mbox{MeV},
\label{eq:fds}
\ee
where the errors indicatively account for all the sources of uncertainty.

The value obtained for $f_{D_s}$ is in reasonable agreement with the
data~\cite{nippe}, $f_{D_s}= 250 \pm 30$ MeV,
and with the lattice results $f_{D_s}=218^{+20}_{-14}$ MeV (quenched),
$235^{+22+17}_{-15-9}$ MeV (unquenched)~\cite{fdslat}, although
within large experimental and theoretical uncertainties.
Our prediction
for  $f_{D_s^*}$ is   in good agreement
with the quenched lattice determination, $f_{D_s^*}=240\pm 20$ MeV~\cite{fdsslat}.

\begin{table}
\centering
\begin{tabular}{|c|rr|rr|}
\cline{2-5}
\multicolumn{1}{c|}{~} &
\multicolumn{2}{c|}{LINSR} & \multicolumn{2}{c|}{NRSX}\\
\hline
MeV  & semileptonic & nonleptonic & semileptonic &
nonleptonic\\
\hline
$f_{D_s}$ &
$304\pm 42\pm 47$ & $253\pm 24\pm 35$ & $297\pm 41\pm 26$ &
$267\pm 25\pm 21$\\
$f_{D_s^*}$ &
$277\pm 36\pm 43$ & $250\pm 31\pm 16$ & $274\pm 36\pm 24$ &
$261\pm 32\pm 9$\\
\hline
\end{tabular}
\caption{{\it Decay constants extracted from both semileptonic
and non-leptonic ratios, eqs.~(\protect\ref{eq:slr}) and
(\protect\ref{eq:nlr}). Both LINSR and NRSX results are shown.
The first error comes from the experimental ones on the BRs, while the
second is a ``theoretical'' error obtained by varying the FP in a range
corresponding to
values of $\chi^2/dof$ up to three times larger than the minimum one.}}
\label{tab:costdec}
\end{table}

\begin{figure}[t]
\centering
\begin{minipage}  {0.05\linewidth}
\vspace*{-10.0cm}
$\delta f_{D_s}$
\end{minipage}
\epsfig{file=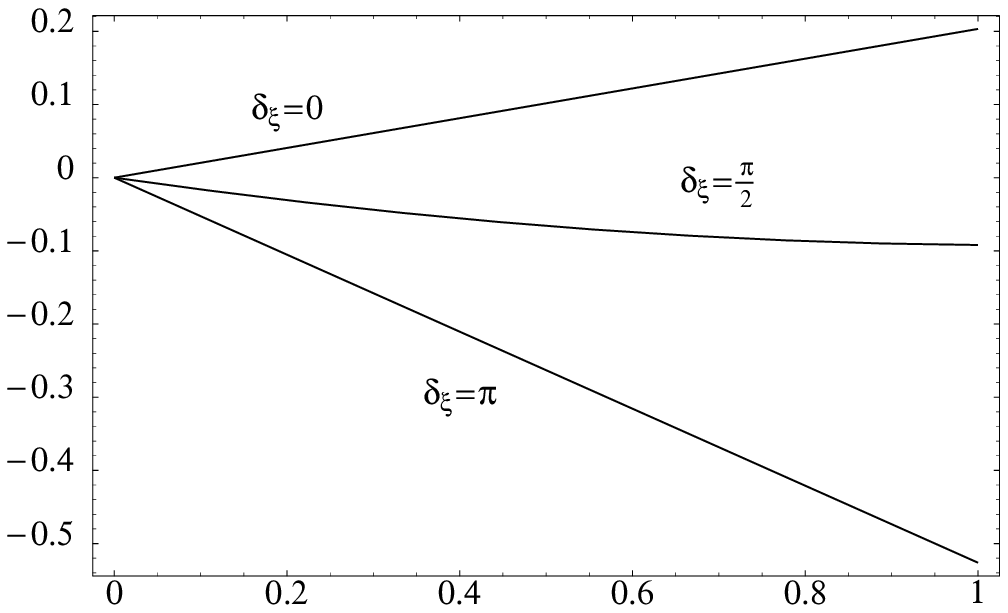,width=0.6\linewidth} \\[5pt]
\hspace*{1.6cm}$\eta_L$
\caption{{\it Charming penguin effect on the determination of
$f_{D_s}$ from semileptonic ratios using NRSX.
The difference $\delta f_{D_s}=f_{D_s}(\eta_L,
\delta_L)/f_{D_s}(0,0)-1$ is plotted as a function of
$\eta_L$ at different values of $\delta_L$. }}
\label{fig:fdsetal}
\end{figure}

Comparing the NRSX results of tab.~\ref{tab:costdec} with the analysis of
ref.~\cite{neuste}, one finds differences of the order of $10$--$15\%$.
Besides our inclusion of the charged decay modes, this difference arises because 
we take into account contributions from penguin operators, which were neglected
in ref.~\cite{neuste}. These contributions amount up to $20\%$ in some channels,
in particular to those used to determine
$f_{D_s}$. For this reason,  it is worth testing the effect of charming-penguin contractions
in the determination of the leptonic decay constants.
We parameterize the effects of charming penguins as in ref.~\cite{chp}, by using two
real quantities $\eta_L$ and $\delta_L$,  denoting the relative size and the 
phase of the charming-penguin amplitudes with respect to the
corresponding emission ones. In fig.~\ref{fig:fdsetal}, we plot
$\delta f_{D_s}=f_{D_s}(\eta_L, \delta_L)/f_{D_s}(0,0)-1$
as a function of $\eta_L$ for various choices of $\delta_L$,
using NRSX.
For $\eta_L\sim 0.2$--$0.3$ and $\delta_L\sim\pi$, as suggested by
$B\to K\pi$ decays~\cite{chp}, $\vert \delta f_{D_s}\vert $ is about $20\%$, larger
than the (previously) estimated theoretical error on $f_{D_s}$. Of course, there is no compelling
theoretical reason to use parameters extracted from $K\pi$ modes
in this analysis. This exercise shows, however, that
penguin effects are not negligible and should be included, at least
as further source of theoretical uncertainty, at the level of 10\%, in
addition to the one in eq.~(\ref{eq:fds}).  
 
\section{Conclusions}
\label{sec:conclusions}

In this paper, we have introduced a parameterization of the 
hadronic matrix elements, that generalizes the one of ref.~\cite{chp} and
express the amplitudes relevant to the calculation of
Cabibbo-allowed non-leptonic $B$ decays in terms of factorized matrix elements
and three real parameters $\alpha$, $\xi$ and $\delta_\xi$.
We have shown the connection of our parameterization
with the generalized factorization of ref.~\cite{neuste}.
In order to fix these parameters, we have reanalized $B\to D\pi$ and
$B\to KJ/\Psi$ data. 

We have found that the fit to $B\to D\pi$ decays
substantially depends on the model describing the Isgur-Wise function $\xi$ and
on the value of its slope. This dependence has been drastically reduced
by fitting the ratios in eq.~(\ref{eq:slr}). We have shown that, once
that the  FP  are  fixed in this way,
a best fit to the non-leptonic $BR(B\to D\pi)$  determines the
values of heavy-heavy form factors at $q^2=M_\pi^2\sim 0$. This provides
a costraint on the HQET models which are currently used for the heavy-heavy
form factors.
We have shown that, in general, different models require different
values of $\hat\rho^2$ to reproduce the fitted values of the form factors.
Consequently, a meaningful test of factorization
is only provided by the comparison  of the values of the form factors
extracted from non-leptonic decays with those directly measured, at small
values of $q^2$, in semileptonic decays. 
The only form factor directly measured at $q^2=0$, $f_+(0)=0.66 \pm 0.06\pm
0.04$~\cite{cleofp}, is in good agreement with our finding, suggesting that
the generalized factorization works well in the case of $B\to D\pi$ decays.

Our best determination of the FP can be found in
(\ref{eq:fitres}), where an estimate of the theoretical
uncertainties is also given. Using these FP, we have also presented a set of 
predictions for the $BR$s of yet-unmeasured $B$ decays, including $D\pi$,  $D\rho$,
$DK^{(*)}$ modes, see tab.~\ref{tab:pred}.

Finally, using  non-leptonic ratios of
eqs.~(\ref{eq:slr}) and (\ref{eq:nlr}), we have extracted the  charmed meson decay
constants from the $BR(B\to DD_s)$, finding
\be
f_{D_s}=270\pm 45~\mbox{MeV},\qquad f_{D_s^*}=260\pm 40~\mbox{MeV},
\ee
where errors indicatively account for all sources of uncertainty present
in the fit, including charming penguins. 

\section*{Acknowledgements}
We warmly thank L.~Lellouch for very useful discussions.
M.~Artuso kindly informed us about the current status of CLEO analyses of
the $B\to D^{(*)}$ form factors. We acknowledge
helpful comments on the experimental results by K.-C. Yang.
M.C., E.F. and G.M.  thank the CERN TH Division
for the hospitality during the completion of this work.
We acknowledge the partial support of the MURST.

\end{document}